\documentclass[aps,prd,a4paper,onecolumn,amsmath,showpacs,superscriptaddress,nofootinbib,preprintnumbers,notitlepage]{revtex4-1}
\usepackage{amssymb,amsmath}
\usepackage{graphicx}
\usepackage{color}
\usepackage{pgfplots}
\usepackage{booktabs}
\usepackage{multirow}
\usepackage{dcolumn}
\usepackage{amsmath}
\usepackage{mathtools}
\usepackage{amsfonts}
\usepackage{amssymb}
\usepackage{epstopdf}
\usepackage{bm}
\usepackage{siunitx}
\usepackage{braket}
\usepackage{enumitem}
\usepackage{soul}
\usepackage{ulem}
\usepackage{color}
\usepackage{transparent}
\usepackage{pifont}
\usepackage[font={small}]{caption}
\usepackage{float}

\usepackage{mathrsfs}
\begin{document}
\title{Reconstructing inflation and reheating in the framework of a generalized $\mathcal{F}(H)$ Friedmann equation}

\author{Ramón Herrera}
\email{ramon.herrera@pucv.cl}
\affiliation{Instituto de F\'{\i}sica, Pontificia Universidad Cat\'{o}lica de Valpara\'{\i}so, Avenida Brasil 2950, Casilla 4059, Valpara\'{\i}so, Chile.
}

\author{Carlos  Ríos}
\email{carlos.rios@ucn.cl}
\affiliation{Departamento de Ense\~nanza de las Ciencias B\'asicas, Universidad Cat\'olica del Norte, Larrondo 1281, Coquimbo, Chile.}

\begin{abstract}
The reconstruction of an inflationary universe considering the parametrization of the scalar spectral index as a function of the number of $e-$folds in the framework of a modified Friedmann equation is analyzed. In this context, we examine the possibility of reconstructing the Hubble parameter together with the effective potential considering a modified Friedmann equation specified by $\mathcal{F}(H)\propto \rho$, where $\mathcal{F}(H)$ corresponds to an arbitrary function of the Hubble parameter $H$ and $\rho$ denotes the energy density associated with the matter in the universe. To reconstruct the background variables during the inflationary scenario, we develop a new methodology by expressing the spectral index in terms of the Hubble parameter and its derivatives. Thus, we obtain a general formalism for the reconstruction of the inflation, using the slow roll approximation together with the parametrization of the scalar spectral index as a function of the number of $e-$folds $N$. As specific examples, we consider the simplest attractor $n_s-1=-2/N$ together with  different functions $\mathcal{F}(H)$, associated to the modified Friedmann equation, to rebuild the Hubble parameter and the effective potential in terms of the scalar field $\phi$. Additionally, we examine the reheating epoch by considering a constant equation of state parameter, in which  we determine the temperature and the number of e-folds during this epoch, using  the background variables found during the reconstruction of the different $\mathcal{F}(H)-$models studied. Besides, we constrain the different parameters associated with the reconstructed inflationary $\mathcal{F}(H)-$models during the epochs of inflation and reheating, using current astronomical data from Planck and BICEP/Keck results.
\end{abstract}

\maketitle
\section{Introduction}

It is widely recognized
that during the early universe, the introduction of the inflationary stage, or inflation, remains a possible solution to many long-standing problems of the hot big bang model see e.g., Refs.\cite{1a,2a,Linde:1990flp}. However, the most significant characteristic of inflation is that it provides a causal explanation for the origin of the observed anisotropy of the cosmic microwave background (CMB) radiation and the distribution of the large-scale structure observed today \cite{6a,7a,8a}.

To describe the inflationary era during the early universe, various inflationary models have been proposed in the context of the theory of general relativity (GR) as well as in modified theories of gravity   or alternatives  to Einstein’s general relativity. The implementation of inflationary models is based   on the introduction of a homogeneous scalar field  $\phi$ associated to the matter of the universe. The  evolution of this scalar field is governed by  the Klein-Gordon equation and together  with the Friedmann equation, and they constitute  the simplest set of field equations utilized  to study the inflationary dynamics of the early universe in the framework of a Friedmann Roberson Walker (FRW) metric. In relation to the modified gravity, we can distinguish  those  models that utilize a modified Friedmann equation to describe the early universe. In this sense, considering a spatially flat FRW metric, the modified Friedmann equation  can be written as
\begin{equation}
\label{F(H)}
    \mathcal{F}(H)=\frac{\kappa}{3}\rho,
\end{equation}
where $\mathcal{F}(H)> 0$ is an arbitrary function associated to the Hubble parameter defined by  $H=(da/dt)/a$, where $a(t)$ denotes  the scale factor. Besides, the quantity $\kappa=8\pi G=M_p^{-2}$ where $M_p$ represents the Planck mass and $\rho$ corresponds to the energy density relates to the matter of the universe. Thus, we have that the dimension associated to the function $\mathcal{F}(H)$ corresponds to  $M_p^2=\kappa^{-1}$. Also, 
we note that   Eq.(\ref{F(H)}) is reduced to the standard Friedmann equation  when the function  $\mathcal{F}(H)=H^2$.

The motivation for considering this modification of the Friedmann equation, defined by Eq.(\ref{F(H)}) arises from the fact that various models in the literature, have analyzed this modification  to describe the early and present universe. 
Thus,  some examples are the Friedmann equations in one anti de Sitter bulk with a Gauss Bonnet (GB) term, in which we have three regimes for the history of the brane universe compacted from the function $\mathcal{F}(H)\propto H^\beta$ \cite{Davis:2002gn,Charmousis:2002rc,Lidsey:2003sj,Herrera:2011zz}. In this case, when we consider the value $\beta=3$, we have the GB regime and then the GB term dominates gravity at highest energies. The situation in which we have $\beta=1$ corresponds to 
 the high energy limit of the brane world cosmology, Randall-Sundrum (RS) regime\cite{Randall:1999ee} and the case $\beta=2$ is the standard Friedmann equation. Besides,  this class of function of the form  power-law type associated to the Hubble parameter  is found   from  the entropy considerations as Tsallis Entopic Proposal or Fractional Entropy (see Refs. \cite{Keskin:2023ngx,Coker:2023yxr}), in which we have $\mathcal{F}(H)\propto H^\beta$.
 
 In the framework of the deformed Ho$\breve{r}$ava-Lifshitz gravity from entropic force, we have that the  function   $\mathcal{F}(H)$ is given by $\mathcal{F}=H^2+\alpha H^4[3-2\ln(4\pi M_p^2/H^2)]$, where   the inverse of the  parameter  $\alpha$ corresponds to  the parameter of Ho$\breve{r}$ava-Lifshitz \cite{Wei:2010wwa}. Besides, corrections to the Friedmann equation inspired by Kaniadakis entropy in which adopting the thermodynamics-gravity conjecture was obtained in Ref.\cite{Sheykhi:2023aqa}. Here the function $\mathcal{F}(H)$ becomes $\mathcal{F}=H^2-\gamma H^{-2}$, where $\gamma$ (with dimensions of $M_p^4$) is associated to Kaniadakis parameter\cite{Lymperis:2021qty}, and the Kaniadakis entropy (or K-entropy), is  one-parameter extension of the classical Boltzmann–Gibbs–Shannon entropy. It emerges from a coherent and self-consistent relativistic statistical framework, maintaining the fundamental aspects of standard statistical theory while reclaiming it under specific conditions\cite{Kaniadakis:2002zz}. Another type of function $\mathcal{F}(H)$ found in the literature corresponds to $\mathcal{F}(H)=H^2+\alpha H^4$, where $\alpha$ is an arbitrary parameter with dimension of $M_{p}^{-2}$. Here the flat Friedmann equation can be derived by considering the Clausius
relation to the apparent horizon of FRW universe, in which entropy is defined to be
proportional to its horizon area plus a logarithmic correction associated with this area\cite{Cai:2008ys}, see also Ref.\cite{Kaul:2000kf}. Also, this function $\mathcal{F}(H)=H^2+\alpha H^4$ or modified Friedmann equation arises when considering an AdS-Schwarzschild black hole via holographic renormalization, incorporating mixed boundary conditions that correspond to the Einstein field equations in four dimensions \cite{Apostolopoulos:2008ru}. Additionally this modified Friedmann equation can be found in the framework of a Chern-Simons type of theory \cite{Gomez:2011zzd,Gomez:2021opp}, in which the parameter $\alpha$ can be positive or negative\cite{Izaurieta:2009hz}. In this sense, there are several other functions  $\mathcal{F}(H)$ or modifications to the Friedmann equation that can be found in the literature, see e.g.,  Refs.\cite{Sheykhi:2010wm,Cai:2008ys,Chen:2008sn,delCampo:2012qb}.

On the other hand, in the literature several authors have analyzed the reconstruction of the background variables, particularly the effective potential associated with a scalar field in the context of inflation, using observational data,  such as, the scalar power spectrum, scalar spectral index $n_s$,  and the tensor to scalar ratio 
$r$\cite{Hodges:1990bf,mas,Herrera:2015udk,Belinchon:2019nhg}.
In this respect, an attractive approach to reconstruct the effective potential of the scalar field using the slow roll approximation, is to consider the parametrization of the observational parameters  in terms of the number of $e$-folds $N$. Specifically, by utilizing the scalar spectral index as a function of $N$ i.e., $n_s=n_s(N)$, often referred to as an attractor, it is possible to rebuild the effective potential as a function of the scalar field \cite{Chiba:2015zpa}. In this context, the simplest  attractor given  by $n_s-1=-2/N$ for large-$N$ (with $N\sim\mathcal{O}(10)\sim\mathcal{O}(10^2)$), aligns with the Planck data, when the number of $e$-folds $N$ is set to 
$N=60$ \cite{BICEP:2021xfz}. By assuming this attractor in the framework of the GR, it is feasible to reconstruct an effective potential\cite{Kallosh:2013hoa} with different limits\cite{1a,Kallosh:2013maa,Linde:1983gd}. In relation to warm inflation, the reconstruction of the effective potential and the dissipation coefficient in terms of the scalar field, was  necessary to consider two attractors; the scalar spectral index and the tensor to scalar ratio in terms of the number of $e-$folds $N$ \cite{Herrera:2018cgi}. Here in the weak dissipative regime, considering the attractors $n_s-1\propto N^{-1}$ together with $r\propto N^{-2}$, it was found that the reconstruction of the effective potential and the dissipation coefficient as functions of the scalar field depends on hyperbolic functions. During the strong regime was obtained that the potential and the dissipation coefficient as functions of the scalar field exhibit a power-law behavior under certain conditions\cite{Herrera:2018cgi}.
Analogously, for the construction of the background variables associated with  Galilean inflation or G-inflation was necessary to utilize two observational parameters \cite{Herrera:2018mvo}. 

In addition, different analysis for  the reconstruction of the background variables using  another parametrizations in terms of the number of $e-$folds $N$ in the context of the slow roll approximation can be found in the literature. Thus, for example, we have the reconstruction of the effective potential as a function  of the scalar field using the parametrization on the slow roll parameter $\epsilon=\epsilon(N)$ \cite{P3, Es2,Es3}. 
Besides, considering the two slow roll parameters $\epsilon(N)$ and $\eta(N)$ as a function of the number of $e-$folds $N$, it is possible  to rebuild the effective potential and the tensor to scalar ratio in terms of the scalar field\cite{EN1,EN2}. For a review of other methodologies to rebuild  the background variables during the inflationary stage, see Refs.\cite{Es8,Es9,Es10,Herrera:2022kes,OTROS1,OTROS2,OTROS3,Gonzalez-Espinoza:2021qnv}. 

On the other hand, at the end of the inflationary stage, the universe undergoes a reheating phase to connect with the standard big bang model \cite{St1,Linde:1990flp}. During this reheating scenario, matter and radiation are produced through the decay of the inflaton field or other fields. As a result, the temperature of the universe rises, eventually leading to the radiation-dominated era and then connecting with the standard hot big bang model. To explain the reheating scenario during  the early universe, various  reheating models are employed to increase the temperature in  this period. One such mechanism involves the perturbative decay of the scalar field through an oscillatory process at the minimum of the effective potential after   the end of inflation \cite{R1a}. Additionally, there are mechanisms associated with non-perturbative analysis, such as;  the parametric resonance decay of the inflaton, the instant reheating model \cite{R4a} or  another field \cite{R2a}. Furthermore, some inflationary models do not involve the inflaton field oscillating around the minimum of the potential; these are known as non-oscillating models (NO models). In such cases, reheating occurs through the decay of another field known as the curvaton \cite{R5a,yo2a}. For further details, see also Ref.\cite{R3} for the reheating from tachyonic instability and other reheating models in Ref.\cite{OTROS5a}.

During the reheating scenario, several important parameters characterize this stage, including the reheating temperature $T_\text{reh}$, the equation of state (EoS) defined as  $\omega_\text{reh}=p/\rho$, which describes the matter content during the reheating era, and the duration of this stage, characterized by the number of 
$e$-folds $N_\text{reh}$. Here, the quantities 
$p$ and $\rho$
 denote the pressure and energy density of the matter-associated fluid during the reheating scenario. We mention that in the case of the reheating temperature,  there is a lower bound imposed by the primordial nucleosynthesis (BBN), given by $T_\text{BBN}\sim 10$ MeV\cite{El1a}. On the other hand, with regard to the EoS parameter, various numerical calculations have been developed to analyze its dynamics based on specific interactions involving the inflaton and other fields related to matter\cite{Podolsky:2005bw}. In this sense, the dynamics evolution in the cosmological time of the EoS parameter depends on the interaction of the inflaton and the another fields. In particular, the authors of Ref.\cite{Felder:2000hq}  determined  that numerically, the EoS parameter exhibited  a slow increase from a value of $\omega_\text{reh}=0$ at the end of inflation to $\omega_\text{reh}\sim 0.3$ during the reheating epoch. In the case of a massive field, it was found that numerically, the EoS parameter increases from a negative value at the end of inflation in which $\omega_\text{reh}=-1/3$, to a value $\omega_\text{reh}\simeq 0$ \cite{Dodelson:2003vq,Martin:2010kz}.
In this sense, as a first approximation and in order to find analytical expressions to the reheating parameters, such as, the temperature $T_\text{reh}$ and the number $N_\text{reh}$, we can  consider that the EoS parameter 
  during  the reheating stage remains approximately constant throughout the this epoch\cite{Munoz:2014eqa}.

The goal of this study is to rebuild inflation in the framework of the generalized Friedmann equation, through the parametrization of an observational parameter in terms of the number of $e-$folds $N$. In particular, we will consider 
an attractor from 
the 
parametrization of the  scalar spectral index $n_s$ as a function of the number of $e-$folds $N$, i.e., $n_s=n_s(N)$.
Here we will develop a new methodology based on rewriting the spectral index in relation to the Hubble parameter and its derivatives.
In this, sense, we study how using different modified Friedmann equations, these affect  the reconstruction of the Hubble parameter together with the effective potential, in terms of the scalar field. We will also establish a general methodology in the context of the slow roll approximation to build the Hubble parameter and the effective potential, by considering  as attractor the scalar spectral index $n_s=n_s(N)$. In this form, settling on a specific scalar spectral index $n_s(N)$, we will    determine the possibility of rebuilding the Hubble parameter as well as the  potential as a function of the scalar field considering different  modified Friedmann equations thought various functions  $\mathcal{F}(H)$.

Besides, we will study the reheating era 
and how the parameters associated to this period, such that, the reheating temperature and number of $e-$folds
  are changed from the reconstruction   of the background variables obtained in the inflationary period (under the slow roll approximation). Thus, 
from these reheating parameters, we will analyze  the reheating temperature and the duration of the reheating  in terms of  the observational parameter $n_s$, from Planck data. In addition,   we will determine how these reheating quantities are constrained considering  different EoS parameters $\omega_\text{reh}$ on the plane $T_\text{reh}=T_\text{reh}(n_s)$ and $N_\text{reh}=N_\text{reh}(n_s)$, respectively.

We organize our paper as follows: In Section \ref{I} we give a brief analyze of the inflationary phase in a generalized Friedmann equation $\mathcal{F}(H)$.
Thus, in this section we show the basic equations under the slow roll approximation during the inflationary epoch to rebuild the background variables. Beside, we present  the observational parameters such as the scalar spectral index, power spectrum together with the tensor to scalar ratio  in this generalized Friedmann equation. 
Additionally,  we obtain under a general formalism, an expression for the Hubble parameter (differential equation)  
in terms  of the number of $e-$folds $N$ to find the reconstruction from any parametrization related to the scalar spectral index $n_s(N)$. 
 In Section \ref{rehet1}, we study the reheating stage under a general formalism in the framework of the modified Friedmann equation. In this section, we express for any function $\mathcal{F}(H)$, the reheating temperature $T_\text{reh}$ together with the number of $e-$folds  $N_\text{reh}$ during this  era.

 In Section \ref{ejep1}, we assume a specific attractor for the scalar spectral index $n_s=n_s(N)$ given by $n_s=1-2/N$, in order to rebuild both the Hubble parameter $H(\phi)$ as the effective potential $V(\phi)$ in terms of the scalar field $\phi$. In this sense, we consider different functions $\mathcal{F}(H)$ to reconstruct the inflationary scenario and the reheating epoch. During the inflationary scenario, we find the different constrains on the parameter-space  from    Planck data.  
In relation to  the reheating scenario, we determine  the reheating temperature together with the number of $e-$folds in this epoch. In this way,  we find these quantities on the  plane $T_\text{reh}=T_\text{reh}(n_s)$ and $N_\text{reh}=N_\text{reh}(n_s)$ for various values of the EoS parameters to constraint the  parameters of our modified model. Finally, in Section \ref{con} we give our conclusions. We chose units in which  $c=\hbar=1$.

\section{ Reconstructing Inflation in a modified Friedmann Equation} \label{I} 

In this section, we will present a brief analysis of the  
 implications of considering a modified Friedman equation characterized through   function 
 $\mathcal{F}(H)
 $ associated with  the Hubble parameter $H$. To describe the matter, in the generalized Friedmann equation $\mathcal{F}(H)
 \propto \rho$, we introduce  that the energy density $\rho$ is associated  to the inflaton field $\phi$. In this way, we can write that the energy density associated to scalar field becomes 
\begin{equation}
\label{rhocampo}
    \rho=\frac{1}{2}\dot{\phi}^2+V(\phi),
\end{equation}
where $V(\phi)$ represents the effective potential related to the inflaton field. Besides, the pressure related to the scalar field is
defined  by 
\begin{equation}
\label{pcampo}
p=\frac{1}{2}\dot{\phi}^2-V(\phi).
\end{equation}
Here we have considered that the scalar field $\phi$ is a homogeneous scalar field i.e., $\phi=\phi(t)$. Also, in the following the dots mean derivatives with respect to the time.

Further, the continuity equation can be written as
\begin{equation}
\label{continuity}
    \dot{\rho}+3H(\rho+p)=0 \,,
\end{equation}
and  replacing   equations (\ref{rhocampo}) and (\ref{pcampo}) in Eq.(\ref{continuity}), we obtain that the dynamics of the scalar field can be written as
\begin{equation}
\label{KleinGordon}
    \ddot{\phi}+3H\dot{\phi}+V_\phi=0,
\end{equation}
where $V_\phi$ represents the derivative of $V(\phi)$ with respect to the $\phi$ field, i.e., $V_\phi=\partial V/\partial \phi$. Besides, in the following, we will use that the notation $\mathcal{F}_H$ corresponds to 
$\mathcal{F}_H=d\mathcal{F}/dH$, $\mathcal{F}_{HH}$ denotes  $\mathcal{F}_{HH}=d^2\mathcal{F}/dH^2$, $H_\phi=dH/d\phi$, $H_{\phi\phi}=d^2H/d\phi^2$, etc.

From Eqs.(\ref{F(H)}), (\ref{rhocampo}) (\ref{pcampo}) and (\ref{KleinGordon}), we find that the speed of the scalar field $\dot{\phi}$
results
\begin{equation}
\dot{\phi}=-\frac{\mathcal{F}_H}{\kappa}\,\left(\frac{H_\phi}{H}\right).\label{df1}
\end{equation}

On the other hand, introducing the number of $e$-folds $N$, provides a way to quantify the amount of inflation required during the expansion of the universe during the inflationary stage. Thus, the number of $e$-folds $N$ defined between two different values of the time; $t$ (or scalar field $\phi$) and $t_\text{end}$ becomes
\begin{equation}
\label{deltaN}
    \Delta N=N-N_\text{end}=\ln\left[\frac{a(t_\text{end})}{a(t)}\right]=\int^{t_\text{end}}_t H\,dt=\kappa\int^{\phi_\text{end}}_\phi\left(\frac{H}{\dot{\phi}}\right)d\phi,
\end{equation}
where  $N_\text{end}$ and $t_\text{end}$ denote the number of  $e$-folds and the time at the end of the inflationary era.

On the other hand,  introducing the   parameters $\epsilon_H $ and $\eta_H $, in the context of a theory with a generalized Friedmann equation we have \cite{delCampo:2012qb}
\begin{equation}
\label{epsilonH1}
    \epsilon_H =-\frac{\dot{H}}{H^2}=-\frac{d\ln H}{d\ln a}= \frac{1}{\kappa}\frac{\mathcal{F}_H}{H}\left(\frac{\,\, H_\phi}{H}\right)^2,\,\,\,\mbox{and}\,\,\,\,\, 
    \eta_H =-\frac{\ddot{H}}{H\dot{H}}= -\frac{d\ln H_\phi}{d\ln a}=\frac{1}{\kappa}\frac{\mathcal{F}_H}{H}\frac{\,\, H_{\phi\phi}}{H}.
\end{equation}
Here we have used the equation for the speed of the scalar field given by Eq.(\ref{df1}).

We note that these   parameters coincide with  the parameters given by the standard slow roll parameters  $\epsilon$ and $\eta$, when $\mathcal{F}(H)=H^2$, under the slow roll approximation. In this way, during slow roll approximation we have that the quantities $\epsilon_{SR}$ and $\eta_{SR}$ when $\mathcal{F}(H)=H^2$ are reduced to
\begin{equation}
    \epsilon_{SR}=\epsilon_H\simeq\frac{1}{2\kappa}\left(\frac{V_\phi}{V}\right)^2,\,\,\,\,\eta_{SR}\simeq\frac{1}{\kappa}\left(\frac{V_{\phi\phi}}{V}\right),\,\,\,\,\,\,\,\mbox{and}\,\,\,\,\,\,\eta_H=\eta_{SR}-\epsilon_{SR}.
\end{equation}
Let us observe that an inflationary scenario of the universe occurs  when $\ddot{a}>0$, which implies that the parameter $\epsilon_H<1$. Furthermore, the end of the inflationary era takes place when $\ddot{a}=0$ or equivalently when the parameter $\epsilon_H=1$.

In relation to the cosmological perturbations, the general perturbed metric about the flat FRW becomes $ds^2=-(1+2A)dt^2+2a(t)B_{,i}dx^{i}dt+a(t)^2[(1-2\psi)\delta_{ij}+2E_{,i,j}+2h_{ij}]dx^{i}dx^{j}$,  where $A$, $B$, $\psi$ and $E$ correspond to the scalar type metric perturbations and the quantity $h_{ij}$ denotes the traverse-traceless tensor type perturbation. In this sense, following Ref.\cite{delCampo:2012qb}, the equation for the Fourier modes  associated to the scalar perturbations can be written as
\begin{equation}
\frac{d^2u_k}{d\eta^2}+\left(k^2-\frac{1}{z}\frac{d^2z}{d\eta^2}\right)u_k=0,\label{M}
\end{equation}
where $u_k$ corresponds to the Mukanov variable defined as $u=z\mathcal{R}$, in which the quantity $z=a \dot{\phi}/H$. Besides,   the variable $\eta$ denotes the conformal time and the quantity  $\mathcal{R}$ is the gauge invariant comovil curvature perturbation. 

The term $(1/z)(d^2z/d\eta^2)$ associated to Eq.(\ref{M}) in the framework of the modified Friedmann equation can be written  as \cite{delCampo:2012qb}
$$
\frac{1}{z}\frac{d^2z}{d\eta^2}=2a^2H^2\Big[1+\frac{1}{2}\epsilon_H\left(5-3H\frac{\mathcal{F}_{HH}}{\mathcal{F}_H}\right)-\frac{3}{2}\eta_H+\frac{1}{2}\eta_H^2
+\epsilon_H^2\Big(1+\frac{H^2}{2}\frac{\mathcal{F}_{HHH}}{\mathcal{F}_H}-H\frac{\mathcal{F}_{HH}}{\mathcal{F}_H}\Big)
$$
\begin{equation}
-\,\,\frac{1}{2}\epsilon_H\,\eta_H\Big(3-2H\frac{\mathcal{F}_{HH}}{\mathcal{F}_H}\Big)\,+\,\mathcal{O}(\dot{\epsilon_H},...).\label{mm}
\end{equation}
In the particular case in which the function $\mathcal{F}(H)=H^2$, the different terms given by Eq.(\ref{mm}) are reduced to those obtained in Ref.\cite{Stewart:1993bc}.

In this form, following the approach outlined in  Ref.\cite{delCampo:2012qb}, the primordial curvature perturbation $A_s$ produced during inflation in the context of a generalized Friedmann equation can be written as 
\begin{equation}
\label{As}
    A_s(k)=\left(\frac{\kappa H^2}{\mathcal{F}_HH_\phi}\right)^2\left(\frac{H}{2\pi}\right)^2.
\end{equation}
 In addition,  the power spectrum of the scalar perturbations $A_s$ given by Eq.(\ref{As}) can be rewritten as
\begin{equation}
\label{AsdeH}
    A_s=\left(\frac{\kappa H^2}{\mathcal{F}_HH_N}\right)\left(\frac{H}{2\pi}\right)^2,
\end{equation}
where we have utilized the relation $\dot{\phi}^2=\mathcal{F}_HH_N/\kappa$.  Here 
the perturbation defined by Eq.(\ref{As}) is  a function of the wave number $k$ and it is evaluated for a particular mode,  when the cosmological scale exits the horizon i.e., when $k=aH$.

By using the primordial curvature perturbation, we can determine the so-called scalar spectral index $n_s$, defined as 
$n_s-1= d\ln A_s/d\ln k$. Thus, from  
Eq.(\ref{As})  the expression for  the scalar spectral index is given by\cite{delCampo:2012qb}
\begin{equation}
\label{nsi}
    n_s-1=2\eta_H-2\left(3-H\frac{\mathcal{F}_{HH}}{\mathcal{F}_H}\right)\epsilon_H.
\end{equation}
We note that in the special case in which $\mathcal{F}(H)=H^2$, Eq.(\ref{nsi}) reduces to the standard expression
$ n_s-1=2\eta_{SR}-6\epsilon_{SR}$.

Additionally, the tensor perturbation (transverse-traceless) during the inflationary scenario would produce gravitational waves, where the tensor perturbations amplitude is denoted by $A_T$. Thus
from these perturbations we can define an important observational quantity called the tensor to scalar ratio $r$, defined as $r=A_T/A_s$. In this form,  we have that the tensor to scalar ratio can be written as \cite{delCampo:2012qb}
\begin{equation}
\label{rdeH}
    r=8\frac{\mathcal{F}_H}{H}\epsilon_H.
\end{equation}

We note that  Eq.(\ref{AsdeH}) and Eq.(\ref{rdeH}) are reduced to the standard expressions  $A_s=\kappa H^2/\left(8\pi^2\epsilon_{SR}\right)$ and $r=16\epsilon_{SR}$ when we choose the function $\mathcal{F}(H)=H^2$ (standard Friedmann equation).

\subsection{Reconstructing  inflation: General reconstruction of $H(\phi)$ and $V(\phi)$ from $n_s(N)$}

In this subsection we consider the methodology  to rebuild the background variables, such as, the Hubble parameter and the effective potential in terms of the scalar field,  assuming as attractor the scalar spectral index in terms of the number of $e$-folds $N$. In this context, we will utilize a new methodology to reconstruct the background variable considering the Hubble parameter and its derivatives in terms of the number of $e-$folds $N$.
 Firstly, we will rewrite the scalar spectral index given by Eq.(\ref{nsi}) in terms  of the Hubble parameter, the function $\mathcal{F}$ and the  derivatives as a function of the number of $e$-folds $N$ through  the parameters $\epsilon_H$ and $\eta_H$, respectively.
 In this way, giving the attractor  $n_s=n_s(N)$, we should first find the Hubble parameter and the effective  potential in terms of the number of $e$-folds $N$. Subsequently, using the  expression given by Eq.(\ref{deltaN}), we should find the $e$-folds $N$ in terms of  the scalar field $\phi$. Finally, considering these equations, we can rebuild the Hubble parameter as a function of the scalar field i.e., $H(\phi)$, and subsequently we can  determine the effective potential $V(\phi)$.

From this methodology, we start by rewriting the parameters $\epsilon_H$ and $\eta_H$ associated to the Hubble parameter in terms of the number of $e-$folds. Thereby, 
we can now rewrite the Hubble parameter and its derivatives in terms of the number $N$ considering  the relation between $N$ and the scalar field from the relation 
\begin{equation}
\label{nphi1}
 dN=-Hdt=-\left(\frac{H}{\dot{\phi}}\right)d\phi, \,\,\,\, \mbox{such that} \,\,\,\, N_\phi=\frac{dN}{d\phi}=-\frac{H}{\dot{\phi}}\,.  
\end{equation}

 In this way, we obtain that
\begin{equation}
\label{Hphi1}
    H_\phi=-\left(\frac{H}{\dot{\phi}}\right)\,H_N.
\end{equation}
Now, using Eq.(\ref{df1}) we find that  the Eq.(\ref{Hphi1}) becomes
\begin{equation}
\label{Hphi2}
    H_\phi^2=\frac{\kappa H^2H_N}{\mathcal{F}_H}.
\end{equation}
Here we note that the ratio $H_N/\mathcal{F}_H$ is a positive quantity.  Besides, as the quantity  $\dot{H}=-H\,H_N<0$, then we have that $H_N>0$ and it suggests that the derivative $\mathcal{F}_H>0$.

Additionally, we can rewrite the quantity $H_{\phi\phi}$ in terms of the derivatives of the number of $e-$folds $N$ as
\begin{equation}
\label{Hphiphi}
    H_{\phi\phi}=\frac{\kappa H^2}{\mathcal{F}_H}\left(\frac{H_{NN}}{2H_N}+\frac{H_N}{H}-\frac{\mathcal{F}_{HH}H_N}{2\mathcal{F}_H}\right).
\end{equation}

Thus, from these relations we can rewrite the parameters $\epsilon_H$ and $\eta_H$, respectively. In this context, 
by replacing Eqs.(\ref{Hphi2}) and (\ref{Hphiphi}) into Eq.(\ref{epsilonH1}), we obtain that these  parameters as a function of $\mathcal{F}$, $H$ and its derivatives with respect to the number $N$ result
\begin{equation}
    \epsilon_H =\frac{H_N}{H}, \,\,\,\,\,\,\mbox{and}\,\,\,\,\,
    \eta_H = \epsilon_H+\frac{H_{NN}}{2H_N}-\frac{\mathcal{F}_{HH}H_N}{2\mathcal{F}_H}.\label{etaH2}
\end{equation}
Using previous results we can rewrite the spectral index $n_s$ given by Eq.(\ref{nsi}) as
\begin{equation}
    n_s-1=\frac{H_{NN}}{H_N}-\frac{4H_N}{H}+\frac{\mathcal{F}_{HH}H_N}{\mathcal{F}_H}=\frac{d\ln H_N}{dN}+g(H)\frac{d\ln H}{dN},  \label{ec1}
\end{equation}
where we have defined the function $g(H)$ as
\begin{equation}
\label{g(H)}
    g(H)=\frac{H\mathcal{F}_{HH}}{\mathcal{F}_H}-4.
\end{equation}

In order to obtain the Hubble parameter  in terms of the number of $e-$folds $N$, we can solve Eq.(\ref{ec1}) to find a first integral given by 
\begin{equation}
\label{eqdiff}
    H_N\exp\left[{G(H)}\right]= \exp\left[{\int(n_s-1)dN}\right],
\end{equation}
where the new function $G(H)$ is given by 
\begin{equation}
\label{G(H)}
    G(H)=\int \frac{g(H)}{H}dH.
\end{equation}

In this way, the Hubble parameter $H=H(N)$ can be obtained from the differential equation given by Eq.(\ref{eqdiff}) assuming a specific attractor for the scalar spectral index $n_s=n_s(N)$.

In order to find the effective potential in terms of the number of $e-$folds i.e., $V(N)$, we can consider the modified Friedmann equation (\ref{F(H)}) obtaining 
\begin{equation}
V(N)\simeq\frac{3}{\kappa}\,\mathcal{F}(H).
\label{Pot}
\end{equation}
Here we have considered the slow roll approximation in which the scalar potential $V\gg\dot{\phi}^2/2$ during the inflationary stage.

Additionally,  to determine a relationship between the number of $e$-folds $N$ and the scalar field $\phi$, we can use the Eqs.(\ref{df1}), (\ref{nphi1}), and (\ref{Hphi2})   to obtain
\begin{equation}
\label{Nphi}
    N_\phi=\sqrt{\frac{\kappa }{\mathcal{F}_H H_N}}\,\,H.
\end{equation}
In this form, integrating Eq.(\ref{Nphi}) we can determine the relation between $N=N(\phi)$ for a specific attractor $n_s=n_s(N)$. Finally, replacing the solution given by Eq.(\ref{Nphi}) into Eq.(\ref{Pot}), we will rebuild the effective potential as a function of the scalar field $\phi$, i.e., $V=V(\phi)$.

\section{ Reheating: General description}\label{rehet1}

In this section, we will analyze the reheating era in the framework of the generalized Friedmann equation in a general description. In this sense, we will utilize the expressions associated to  the background variables to find the reheating parameters, such as, the reheating temperature, as well as, the number of $e-$folds during this epoch.  
 To start with this study, we can assume that the physical scale  cross the horizon during inflation  when the wave number $k$ is equal to  $k=a_k H_k$. In addition, we can assume that the physical scale crosses the horizon at the current epoch when  $k_0=a_0H_0$. Here, the subscript 
``$k$''
 denotes that the quantities are evaluated when $k=a_k H_k$, and the subscript 
``0'' represents the physical quantities evaluated at the present time. Besides, the ratio between the wave numbers $k/k_0$ can be written as 
\begin{equation}
\label{koverk}
	\frac{k}{k_0}=\frac{a_kH_k}{a_0H_0}=\left(\frac{a_k}{a_\text{end}}\right)\left(\frac{a_\text{end}}{a_\text{reh}}\right)\left(\frac{a_\text{reh}}{a_\text{eq}}\right)\left(\frac{a_\text{eq}H_\text{eq}}{a_0H_0}\right)\left(\frac{H_k}{H_\text{eq}}\right).
\end{equation}
As before, here we have used that the ``end'' subscript means that the variable is evaluated at the end of inflation. In addition, the notation  ``reh'' corresponds  to the reheating era and ``eq'' denotes  the   of radiation-matter equality. 

In fact, we can write the number of $e$-folds $N$ in each stage (duration) as a function of the scale factor $a$ as follows   
	
 $$
 N_k=\ln\left(\frac{a_\text{end}}{a_k}\right),\,\,\,\,\,\,	N_\text{reh}=\ln\left(\frac{a_\text{reh}}{a_\text{end}}\right)\,\,\,\,\;\;
 \mbox{and}\,\,\,\,\, \;\;N_{RD}=\ln\left(\frac{a_\text{eq}}{a_\text{reh}}\right),
 $$
respectively. Here, the notation 
 $N_\text{reh}$ corresponds to the number of $e$-folds during reheating era and $N_\text{RD}$ denotes to the number of $e$-folds in the radiation dominance (RD).
By using these different $e-$folds in each era, we rewrite  Eq.(\ref{koverk}) as
\begin{equation}
\label{NewNreh}
	\text{ln}\left(\frac{k}{a_0H_0}\right)=-N_k-N_\text{reh}-N_\text{RD}+\text{ln}\left(\frac{a_\text{eq}H_\text{eq}}{a_0H_0}\right)+\text{ln}\left(\frac{H_k}{H_\text{eq}}\right).
\end{equation}
Besides, we can utilize the EoS parameter  $\omega_\text{reh}$ related  to the reheating regime to express the ratio between the energy density at the end of inflation  $\rho_\text{end}$ and the energy density during the reheating stage $\rho_\text{reh}$. In this way,     the ratio $\rho_\text{reh}/\rho_\text{end}$ can be written as
\begin{equation}
	\frac{\rho_\text{reh}}{\rho_\text{end}}=e^{-3N_\text{reh}(1+\omega_\text{reh})}.\label{rend}
\end{equation}
For this expression 
 we have considered that during the reheating stage the energy density $\rho$ has a dependence   with the scale factor $a$ as; 
 $\rho\propto a^{-3(1+\omega_\text{reh})}$, in which the parameter $\omega_\text{reh}$ is assumed a constant.
 
In order to determine the enrgy density of the field at the end of inflation $\rho_\text{end}$ in Eq.(\ref{rend}), we can  first rewrite the parameter $\epsilon_H$ from  Eq.(\ref{epsilonH1}) as 
\begin{equation}
    \epsilon_H = \frac{2\kappa}{H\mathcal{F}_H}\left(\frac{\,\dot{\phi}^2}{2}\right)=\frac{2\kappa}{H\mathcal{F}_H}\left(\rho-V\right),
\end{equation}
where we have used  Eqs.(\ref{F(H)}) and (\ref{rhocampo}), respectively. Considering that the end of the inflationary era occurs
when $\epsilon_H=1$ (or equivalently $\ddot{a}=0$), we can find from the above equation that the energy density at the end of the inflationary era 
$\rho_\text{end}$ can be obtained from the relation 

\begin{equation}
\label{EqRhoEnd}
    \rho_\text{end}\,-\,\frac{1}{2\kappa}\left.\left(\frac{d\mathcal{F}}{d\ln H}\right)\right|_\text{end}\,=\,V_{\text{end}},
\end{equation}
where we mention that the second term of Eq.(\ref{EqRhoEnd}) is a function of $\rho_\text{end}$ and it depends of the function $\mathcal{F}(H)$. 

In the particular case in which  the function $\mathcal F(H)$ corresponds to the standard Friedmann equation i.e., $\mathcal{F}(H)=H^2$, we have that the term $\left.\left(d\mathcal{F}/d\ln H\right)\right|_\text{end}=(2\kappa/3)\rho_\text{end}$. Thus,  substituting this result into Eq.(\ref{EqRhoEnd}) we find  the well known result for  $\rho_\text{end}$ given by $\rho_\text{end}=(3/2)V_\text{end}$
\cite{Munoz:2014eqa,Dai:2014jja,Cook:2015vqa}. 

On the other hand, to find the reheating temperature $T_\text{reh}$,  we can consider 
the entropy conservation, in which  the  entropy generated during reheating
is preserved in the CMB together with the  neutrino background at the current epoch \cite{Dai:2014jja}. In this context, following Ref.\cite{Dai:2014jja} we can write this conservation as
\begin{equation}
g_{\text{s,reh}}a^3_{\text{reh}}T^3_{\text{reh}}=a_0^3\left(2T_0^3+\frac{21}{4}T_{\nu,0}^3\right),\label{RR1}
\end{equation}
where the quantity $g_{\text{s,reh}}$ corresponds to  the effective number of relativistic degrees of freedom for entropy at reheating, the temperature $T_0$ denotes  the present CMB temperature i.e., $T_0\simeq2.7$K  and the quantity $T_{\nu,0}$ is the present neutrino temperature. Following Ref.\cite{Dai:2014jja} we can consider that the 
 relation between the temperatures  $T_{\nu,0}$ and $T_0$ is given by 
 $T_{\nu,0}=\left(4/11\right)^{1/3}T_0$, and then from Eq.(\ref{RR1})  we can relate  the  scale factors during the reheating scenario and at the current era from the expression
$a_{\text{reh}}/a_0=\left[43/(11g_{\text{s,reh}})\right]^{1/3}T_0/T_{\text{reh}}$.

In addition, we can consider that the energy density at the end of reheating $\rho_{\text{reh}}$ corresponds  to the hot radiation $\rho_{\text{reh}}\propto T^4_{\text{reh}}$, with which we can write
\begin{equation}
\label{Treh1}
	\rho_{\text{reh}}=\frac{\pi^2}{30}g_{\star\text{,reh}}T^4_{\text{reh}}\,,
\end{equation}
where the quantity $g_{\star\text{,reh}}$ represents  the effective number of relativistic degrees of freedom at the end of reheating scenario. 

In this form, utilizing    the above expressions,  we find that the reheating temperature $T_\text{reh}$ in terms of the parameters $\rho_\text{end}$, $\omega_\text{reh}$ and $N_\text{reh}$ becomes
\begin{equation}
\label{Treh}
	T_\text{reh}=\text{exp}\left[-\frac{3}{4}(1+\omega_\text{reh})N_\text{reh}\right]\left[\frac{30\,\rho_\text{end}}{g_{\star\text{,reh}}\pi^2}\right]^{1/4}.
\end{equation}

Besides, we find that  the duration of the reheating epoch characterized by  
the number of $e$-folds  $N_\text{reh}$  results
\begin{equation}
\label{Nreh}
N_\text{reh} =\frac{4}{1-3\omega_\text{reh}}\left[- N_k-\ln\left(\frac{k}{a_0T_0}\right)  -\frac{1}{3}\text{ln} \left(\frac{11g_\text{s,reh}}{43}\right)
	-\frac{1}{4}\ln\left(\frac{30 \kappa^2\rho_\text{end}}{g_{\star\text{,reh}}\pi^2}\right)+\frac{1}{2}\text{ln}\left(\frac{\pi^2 r A_s}{2}\right)\right].
\end{equation}
Here we note that to obtain the energy density at the end of the inflationary scenario $\rho_\text{end}$, we need to solve the Eq.(\ref{EqRhoEnd}) for the different  functions $\mathcal{F}(H)$, and then to write  the density
$\rho_\text{end}$  in terms of the  potential at the end of inflation i.e., 
$\rho_\text{end}=\rho_\text{end}(V_\text{end})$.
In this sense, we have analyzed the reheating scenario for any $\mathcal{F}(H)-$model under one general description. In the following, we will apply our methodology for different  functions $\mathcal{F}(H)$ for the simplest attractor for the scalar spectral index in terms of the number of $e-$folds given by  $n_s=1-2/N$.

\section{Reconstruction from the scalar spectral index $n_s=n_s(N)$ }\label{ejep1}

In this section we will make use of the methodology earlier described, considering as attractor one specific parametrization for the scalar spectral index as a function of the number of $e-$folds $N$. In this context, assuming different functions $\mathcal{F}(H)$, we will   reconstruct the Hubble parameter together with the effective potential  in terms of the scalar field. In addition, in this section we will analyze the reheating era,  considering the background  variables  obtained in the preceding section from of  different functions $\mathcal{F}(H)$.

In order to give a scalar spectral index $n_s$ as a function of the number of $e-$folds $N$, we consider the simplest parametrization (or attractor)   $n_s=n_s(N)$ defined  by \cite{{Chiba:2015zpa}}
\begin{equation}
\label{attractor}
n_s(N)=1-\frac{2}{N},    
\end{equation}
with $N\neq$ 0.
Here as we mentioned before, this parametrization 
 is in agreement with Planck's measurements, when the number of $e-$folds becomes $N\simeq 60$  
 \cite{BICEP:2021xfz}.

In addition, to test the modified Friedmann model, we will assume that  the reconstruction of the background variables will be developed in light of three types of functions $\mathcal{F}(H)$ associated to the Hubble parameter $H$. First, we will consider that the function  $\mathcal{F}(H)$ of the   power-law type $\mathcal{F}(H)=\alpha_\lambda H^{\beta}$. As a second example, we will assume  that  the  function of type $\mathcal{F}(H)=H^2-\gamma H^{-2}$ and finally we will choose   a function of type $\mathcal{F}(H)=H^2\pm\theta H^{4}$ to study the reconstruction and the reheating of these $\mathcal{F}(H)-$models.

\subsection{Reconstruction example I: $\mathcal{F}(H)=\alpha_\lambda H^{\beta}$}\label{RecI}

The first example that we will consider to reconstruct the background variables is the case in which $\mathcal{F}(H)$ is a  function of the power-law type. In this sense, we will analyze  the function defined as
\begin{equation}
\label{F1}
\mathcal{F}(H)=\alpha_\lambda H^\beta,    
\end{equation}
where the parameters  $\alpha_\lambda$ and $\beta$ are positive constants.
We mention that the dimension of the parameter $\alpha_\lambda$ is $[\alpha_\lambda]=M_p^{2-\beta}$ and  $\beta$  is a  dimensionless quantity. In particular, for the standard Friedmann equation in which $\mathcal{F}(H)=H^2$, we have that $\beta=2$ and $\alpha_\lambda=1$. Besides, 
as we mentioned before  this class of function associated to the Hubble parameter $\mathcal{F}(H)\propto H^\beta$ is obtained in different limits in a GB theory, as well  in entropy considerations.

To rebuild the Hubble parameter in terms of the number of $e-$folds $N$, firstly we can replace     
 Eq.(\ref{F1}) into Eq.(\ref{g(H)}) to obtain that the function $g(H)$ becomes
\begin{equation}
  g(H)=\frac{H\mathcal{F}_{HH}}{\mathcal{F}_H}-4=\beta-5=\mbox{constant}.  
\end{equation}
Now, introducing this result into Eq.(\ref{G(H)}) and using Eqs.(\ref{eqdiff}) and  (\ref{attractor}),
we find that the Hubble parameter as a function of the number of $e-$folds $N$ yields
\begin{equation}
\label{HdeN1}
    H(N)=\left[(4-\beta)\left(\frac{A_1}{N}+B_1\right)\right]^{1/(\beta-4)}\,\,\,>0,\,\,\,\,\mbox{with}\,\,\,\,\beta\neq 4,
\end{equation}
and the quantities $A_1$ and $B_1$ are two arbitrary integration constants which have units of $M_p^{\beta-4}$, since the Hubble parameter has dimension of $M_p$. In general these integration constants are positives, negatives o zero. However, in what follows, 
we will assume for simplicity that the integration constant $B_1>0$, while the constant $A_1$ must be positive, as we will see later.

To rebuild the Hubble parameter in terms of the scalar field $H=H(\phi)$, we have  to find the number of $e-$folds $N$ as a function of the scalar field $\phi$ i.e., $N=N(\phi)$.  Thus, from  Eq.(\ref{Nphi}) we get
\begin{equation}
   \frac{dN}{d\phi}= N_\phi=\sqrt{\frac{\kappa}{\alpha_\lambda\beta A_1}}\left(\frac{N}{H}\right)=\sqrt{\frac{\kappa}{\alpha_\lambda\beta A_1}}N\left[(4-\beta)\left(\frac{A_1}{N}+B_1\right)\right]^{-1/(\beta-4)},
\end{equation}
and solving this differential equation, we obtain that the solution for the number  $N(\phi)$ can be written as
\begin{equation}
    N(\phi)=\mathcal{J}^{-1}(\phi),
\end{equation}
where the quantity $\mathcal{J}^{-1}(\phi)$ represents the inverse of the function $\mathcal{J}(\phi)$ defined by
$$
    \mathcal{J}(\phi)=\frac{(4-\beta)^{\left(\frac{\beta-3}{\beta-4}\right)}}{(3-\beta)B_1}\left(B_1+\frac{A_1}{\sqrt{\kappa/(\alpha_\lambda\beta A_1)}\phi+\tilde{c}_1}\right)^{\left(\frac{\beta-3}{\beta-4}\right)}\,\,\,\,\,\,\times
$$  
\begin{equation}  
_2F_1\left[1\,\,,1+\frac{1}{\beta-4},2+\frac{1}{\beta-4},1+\frac{A_1}{B_1(\sqrt{\kappa/(\alpha_\lambda\beta A_1)}\phi+\tilde{c}_1)}\right],\label{Hf}
\end{equation}
where $\tilde{c}_1$ is a new constant of integration, and the function $_2F_1$ corresponds to the hypergeometric function \cite{L20}.

Thus, from Eqs.(\ref{HdeN1}) and (\ref{Hf}), we find that   the reconstruction of the Hubble parameter   as a function of $\phi$ can be written as
\begin{equation}
\label{HdePhi1}
    H(\phi)=\left[(4-\beta)\left(\frac{A_1}{\mathcal{J}^{-1}(\phi)}+B_1\right)\right]^{1/(\beta-4)}.
\end{equation}
In this way, using now the Eq.(\ref{F(H)}) in the slow roll approximation, in which $V\simeq3 \alpha_\lambda H^\beta/\kappa $ (see Eq.(\ref{Pot})), we  obtain that the reconstruction of the  the potential $V(\phi)$ is given by
\begin{equation}
    V(\phi)=\frac{3\alpha_\lambda}{\kappa} \left[(4-\beta)\left(\frac{A_1}{\mathcal{J}^{-1}(\phi)}+B_1\right)\right]^{\beta/(\beta-4)}.
\label{POTE}
\end{equation}
Here we note that in the special case in which the parameter $\beta=2$ and then $\alpha_{\lambda=2}=1$, the reconstruction of the effective potential in terms of the scalar field defined by Eq.(\ref{POTE}) is reduced to standard expressions  obtained in GR (T-model) \cite{Chiba:2015zpa} in which 
\begin{equation}
    V(\phi)=\frac{3}{2 \kappa B_1}\tanh^2\left[\frac{1}{2}\left(\sqrt{\frac{\kappa B_1}{ A_1}}\phi+\bar{c}_1\right)\right].
\end{equation}

On the other hand, in order to constraint the free parameters of our model, we will consider the amplitude of the power spectrum of the scalar perturbations given by Eq.(\ref{AsdeH}) together with the tensor-scalar ratio given by Eq.(\ref{rdeH}). In this form, we have that the amplitude of the power spectrum of the scalar perturbations results
\begin{equation}
\label{Delta(N)}
    A_s=\frac{\kappa}{4\pi^2}\frac{N^2}{(\alpha_\lambda\beta A_1)}\,\,\,\Rightarrow \,\,\,\Lambda=(\alpha_\lambda\beta A_1)=\frac{\kappa}{4\pi^2}\frac{N^2}{A_s}.
\end{equation}
Here we note that since that the parameters $\alpha_\lambda$ and $\beta$ are positives quantities, then from the above expression we have that the integration constant $A_1>0$. If we consider the case in which the number of $e$-folds during the crossing epoch is $N_k=60$ and $A_s=2.2\times 10^{-9}$, we obtain from Eq.(\ref{Delta(N)}) that $A_1\simeq 4.14\times10^{10}M_p^{-2}/\alpha_\lambda \beta$. In particular for the standard Friedmann equation in  which the parameter $\beta=2$ and $\alpha_2=1$, we obtain the constraint on the parameter $A_1\approx2\times10^{10}M_p^{-2}$. 

On the other hand, the tensor to scalar ratio $r$ is calculated as follows
\begin{equation}
\label{r1N}
    r=\frac{8\Lambda H^2}{N^2}=\frac{2\kappa}{\pi^2 A_s}\left[(4-\beta)\left(\frac{A_1}{N}+B_1\right)\right]^{2/(\beta-4)},
\end{equation}
where we have used Eq.(\ref{HdeN1}).
Since that the tensor to scalar ratio $r$ is a real and positive quantity, we can constraint the parameter $\beta$ to  $0<\beta<4$. 

Besides, using Eq.(\ref{r1N}) we will obtain a constraint for the parameter $B_1$. To do this, we consider the fact that $r$ evaluated at the time of crossing is constrained by the value $\left.r(k)\right|_{k=a_kH_k}=r_k<0.039$.  Thus, from this observational parameter we have
\begin{equation}
\label{B1Inequation}
    B_1>\frac{1}{4-\beta}\left(\frac{2\kappa}{\pi^2 A_s r_k }\right)^{\left(4-\beta\right)/2}-\frac{\kappa N_k}{4\pi^2\alpha_\lambda\beta A_s}.
\end{equation}
Since $B_1>0$, we can obtain a lower bound for the parameter $\alpha_\lambda$ given by
\begin{equation}
\label{A1Inequation}
    \alpha_\lambda > \frac{(4-\beta)\kappa N_k}{4\pi^2\beta A_s}\left(\frac{2\kappa}{\pi^2A_s r_k}\right)^{(\beta-4)/2}.
\end{equation}

Additionally, to find the number of $e-$folds at the end of the inflationary epoch $N_\text{end}$, we can consider that the parameter $\epsilon_H(N)$ given by Eq.(\ref{epsilonH1}) can be rewritten as
\begin{equation}
    \epsilon_H=\frac{H_N}{H}=\frac{A_1}{(4-\beta)N^2}\left(\frac{A_1}{N}+B_1\right)^{-1}=\frac{1}{(4-\beta)(1+\mu N)N},
\end{equation}
where we have used Eq.(\ref{HdeN1}), and we have defined the dimensionless parameter  $\mu=B_1/A_1$.
 Thus, the above equation can be evaluated at the end of inflation, where $\epsilon_H=1$ (or equivalently $\ddot{a}=0$). In this way, we  find the number of $e-$folds at the end of the inflationary era $N_\text{end}$ becomes
\begin{equation}
\label{NendFromMu}
N_\text{end}=\frac{\sqrt{1+\left(\frac{4}{4-\beta}\right)\mu}\,-\,1}{2\mu}.
\end{equation}
Here we note that for the special case in which the parameter  $\mu\ll 1$, the number of $e-$folds at the end of inflation tends to zero, i.e., 
$N_\text{end}\sim 0$. While in the situation in which $\mu\gg 1$ the number of $e-$folds $N_\text{end}\sim [(4-\beta)\mu]^{-1/2}$.

\begin{figure}[h]
    \centering
    \includegraphics[width=0.8\linewidth]{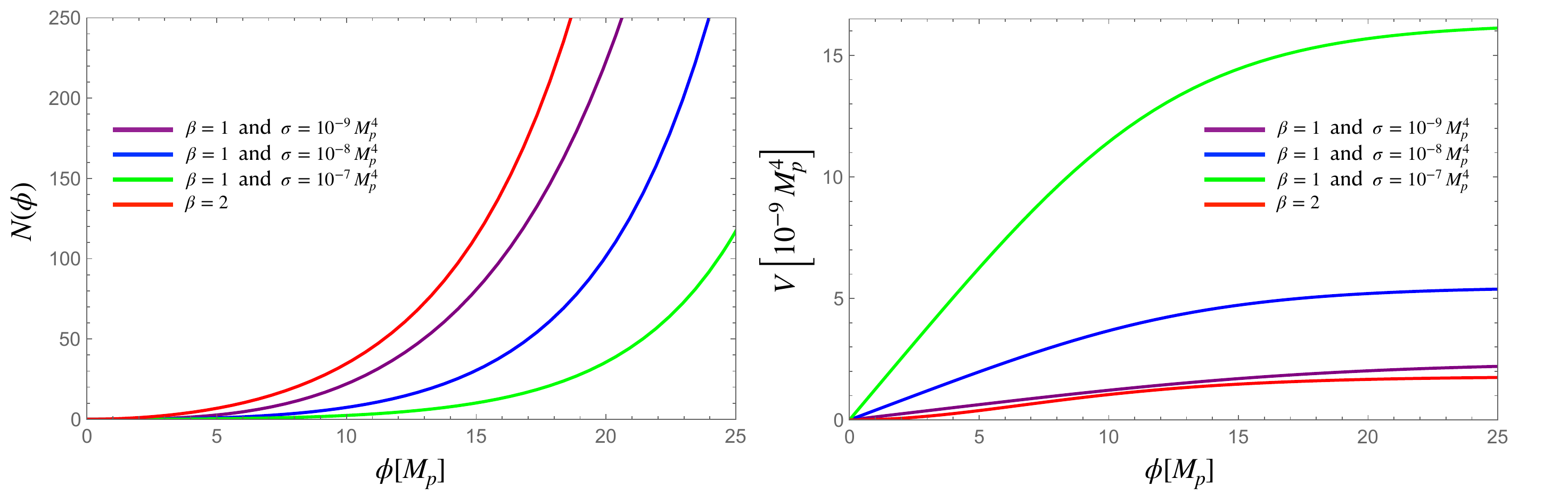}
    \caption{Evolution of the number of $e-$folds (left panel) and the effective potential (right panel) versus the scalar field for two different values of the parameter $\beta$. In particular for the case in which $\beta=1$, we have considered three values on  the brane tension  $\sigma$. Also, in both panels we have used that the integration constant $\bar{c}_1=0$ and the number $N_k=60$. }
    \label{Fig_Potential_M1}
\end{figure}

In Fig.\ref{Fig_Potential_M1} shows the evolution of the number of $e-$folds (left panel) and 
the effective  potential (right panel) as a function of the scalar field for two different values of the parameter $\beta$. In addition, when $\beta=1$ (or $\alpha_{\lambda=1}=1$ ), we show three values for the  brane tension $\sigma$, associated to the parameter $\alpha_{\lambda=1}$ from the relation $\alpha_{\lambda=1}=\left(2\kappa\sigma/3\right)^{1/2}$. In this form, for $\beta=1$, we consider the values; $\sigma=10^{-9}M_p^4$, $\sigma=10^{-8}M_p^4$ and $\sigma=10^{-7}M_p^4$, which give the values $\alpha_1=2.58\times10^{-5}M_p$ (purple curve), $\alpha_1=8.16\times10^{-5}M_p$ (blue curve) and $\alpha_1=2.58\times10^{-4}M_p$ (green curve), respectively. Besides, we have obtained for $\sigma=10^{-9}M_p^4$ the values $A_1=1.61\times10^{15}M_p^{-3}$ and  $B_1=1.15\times10^{13}M_p^{-3}$ associated to the purple curve, for the brane tension $\sigma=10^{-8}M_p^4$ the values $A_1=5.08\times10^{14}M_p^{-3}$ and  $B_1=2.98\times10^{13}M_p^{-3}$  (blue curve) and the values $A_1=1.61\times10^{14}M_p^{-3}$ and $B_1=3.56\times10^{13}M_p^{-3}$ for $\sigma=10^{-7} M_p^4$ (green curve). 
 In addition, we have considered the special case  $\beta=2$ in which $\alpha_2=1$ (red curve), which as mentioned before represents the standard Friedmann equation. For this situation we have found  the values $A_1=2.07\times10^{10}M_p^{-2}$ and  $B_1=8.35\times10^{8}M_p^{-2}$. In these plots we have considered that the constant $\bar{c}_1=0$ for simplicity and the number of $e-$folds $N_k=60$. In relation to the right panel of this figure, we note that the reconstructed effective potential in terms of the scalar field shows a maximum value given by a flat region for large-$\phi$ ($\phi>15 M_p$) in which the number of $e-$folds $N$ is also large ($N>30$), see left panel. In addition, 
 from the right panel, we can note that scalar field begins to roll from the maximum value of the potential (flat region) towards values of the scalar field $\phi\sim 0$ where the number of $e-$folds at the end of inflation $N_\text{end}$ is approximately zero.

\begin{figure}[h]
    \centering  \includegraphics[width=0.35\linewidth]{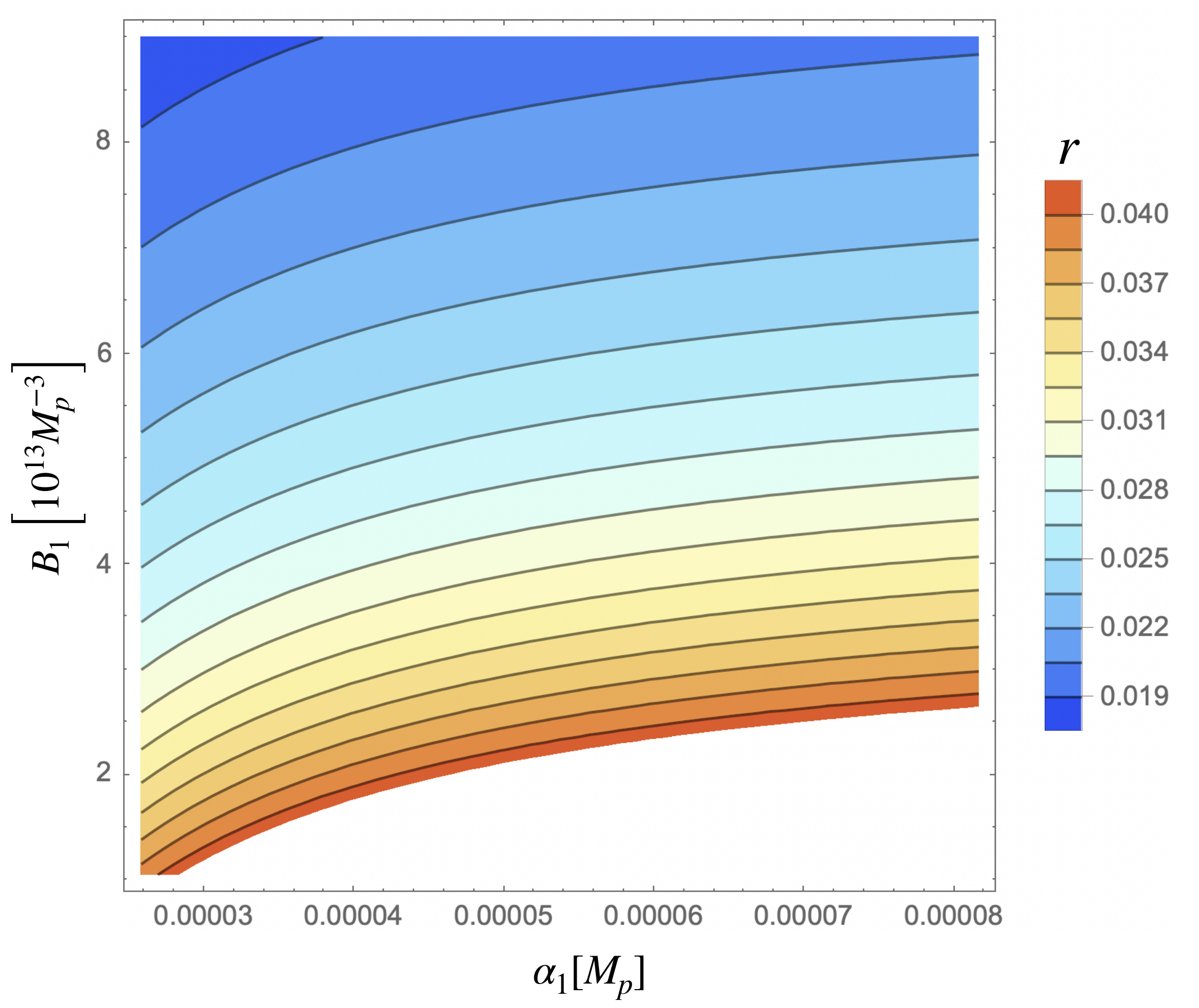}
    \caption{Contour plot for the tensor to scalar ratio $r$ in terms of the integration constant $B_1$ and the parameter associated to  brane tension $\alpha_1$. Here we have fixed the number of $e-$folds $N_k=60$ and $\bar{c}_1=0$.}
    \label{Fig_Contour_M1}
\end{figure}

 In Fig.\ref{Fig_Contour_M1} we show the contours curves  associated to the tensor to scalar ratio $r$, and different combinations of the parameters $\alpha_1$ and $B_1$ from Eq.(\ref{r1N}) for the special case in which $N=60$. From this plot, given a value of the tensor to scalar ratio from the vertical column we can constrain the parameter space of $B_1$-$\alpha_1$.

\begin{figure}[h]
    \centering
    \includegraphics[width=0.6\linewidth]{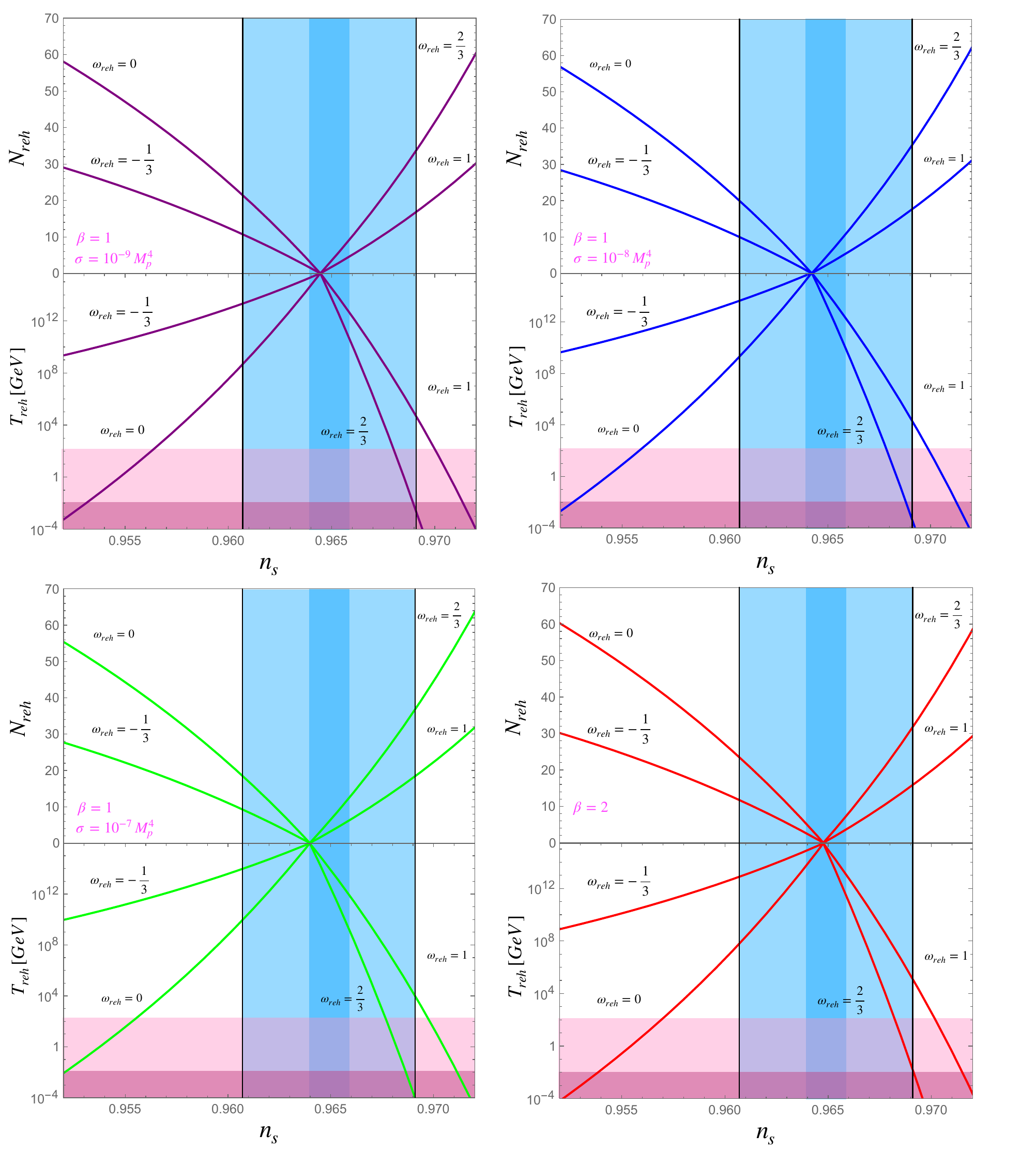}
    \caption{ The figure shows the number of $e-$folds during the reheating scenario $N_\text{reh}$ (upper panels) and the reheating temperature $T_\text{reh}$ (lower panels)  versus the scalar spectral index $n_s$ for different values of the parameter $\beta$. In particular for the special case in which the power $\beta=1$, we have considered three different values for the  brane tension $\sigma$.
     For the case $\beta=1$ and the  brane tension $\sigma=10^{-9}M_p^4$ corresponds to the upper left panel. The case $\beta=1$ and  
    $\sigma=10^{-8}M_p^4$ is for the upper right panel. Again the case $\beta=1$ and 
$\sigma=10^{-7}M_p^4$ corresponds to the lower left panel  and the situation in which $\beta=2$
i.e., the standard Friedmann equation is the lower right panel.        Besides,  in all the panels we have assumed different values of the EoS parameter $\omega_\text{reh}$ associated to the reheating epoch; $\omega_\text{reh}=-1/3,0,2/3$ and $\omega_\text{reh}=1$, respectively. Also, in each of the  panels, we have utilized  the parameters shown in Table I. } 
    \label{Fig_Reh_M1}
\end{figure}

On the other hand, we will study the reheating epoch  using the reconstruction of the background variables derived during the inflationary scenario. To analyze the reheating era, we find that  the energy density at the end of inflation, from Eq.(\ref{EqRhoEnd}) yields

\begin{equation}
\label{rhoend1}
    \rho_\text{end}=\left(\frac{1}{1-\beta/6}\right)V_\text{end}.
\end{equation}
Here as mentioned before, we note that if $\beta=2$, we recover the standard case for the Friedmann equation and the energy density at the end of inflation is reduced to $\rho_\text{end}=(3/2)V_\text{end}$.
By using Eq.(\ref{rhoend1}) and replacing into  Eqs.(\ref{attractor}) and (\ref{r1N}), we find that  the reheating temperature $T_\text{reh}$ and the number of $e-$folds during the reheating $N_\text{reh}$, given by Eqs.(\ref{Treh}) and (\ref{Nreh})  as a function of the spectral index $n_s$ become
\begin{equation}
\label{TrehM1}
	T_\text{reh}(n_s)=\text{exp}\left[-\frac{3}{4}(1+\omega_\text{reh})N_\text{reh}(n_s)\right]\left[\frac{30\,V_\text{end}}{g_{\star\text{,reh}}\pi^2(1-\beta/6)}\right]^{1/4},\,\,\,\mbox{and}
\end{equation}
\begin{equation}
\label{NrehM1}
N_\text{reh}(n_s) =\frac{4}{1-3\omega_\text{reh}}\left[\frac{2}{n_s-1}-\ln\left(\frac{k}{a_0T_0}\right)  -\frac{1}{3}\text{ln} \left(\frac{11g_\text{s,reh}}{43}\right)
	-\frac{1}{4}\ln\left(\frac{30 \kappa^2V_\text{end}}{g_{\star\text{,reh}}\pi^2(1-\beta/6)}\right)+\tilde{r}_1(n_s)\right], 
\end{equation}
where $\omega_\text{reh}\neq 1/3$ and  we have defined the function $\tilde{r}_1(n_s)$ as
\begin{equation}
    \tilde{r}_1(n_s)=\ln\left\{\kappa^{1/2}\left[(4-\beta)\left(\frac{(1-n_s)}{2}A_1+B_1\right)\right]^{1/(\beta-4)}\right\}.
\end{equation}
The Fig.\ref{Fig_Reh_M1} shows the evolution of the temperature $T_\text{reh}(n_s)$ and the number of  $e-$folds $N_\text{reh}(n_s)$ during the reheating epoch  versus the scalar spectral index from Eqs.(\ref{TrehM1}) and (\ref{NrehM1}), respectively. Four graphs have been plotted:  three plots represent different values of the  brane tension $\sigma$ in which  $\beta=1$, while one corresponds to $\beta=2$ reflecting the standard Friedmann equation. In all the panels different values have been used for the EoS parameter $\omega_\text{reh}$ related to the reheating epoch  and these values are; $\omega_\text{reh}=\{-1/3,0,2/3,1\}$. 
In this figure,  the different regions  are; the light blue shaded region
corresponds to  
the maximum and minimum bounds of the scalar spectral index $n_s=0.9649\pm 0.0042$ from  Planck data  at 1$ \sigma$ limits. Also,  the blue shaded denotes to a projected sensitivity from the central value  of the scalar spectral index  ($n_s=0.9649$) with an uncertainly  of $\pm 10^{-3}$ in the test analyzed by the authors in 
Ref.\cite{-3}. Besides, 
the pink shaded region is  the electroweak scale where the temperature   $T_\text{EW}\sim100\text{GeV}$ and the purple shaded region corresponds to a temperatures below   10 MeV,  and this region is disapproved by primordial nucleosynthesis.

In  the upper left panel of Fig.\ref{Fig_Reh_M1}, we have considered the parameter $\beta=1$ in which the Friedmann equation is  $H\propto \rho$, and we have chosen  that the tension of the brane   $\sigma=10^{-9}M_p^4$ (or equivalently $\alpha_1=2.58\times10^{-5}M_p$). Here we have determined that the   values of  the parameters  $A_1$, $B_1$ and $N_\text{end}$ correspond to;  $A_1=1.61\times10^{15}M_p^{-3}$, $B_1=1.15\times10^{13}M_p^{-3}$ and $N_\text{end}=0.33$. 
In  the upper right panel, we have also $\beta=1$ and the for the tension of the brane the value $\sigma=10^{-8}M_p^4$ ( or $\alpha_1=8.16\times10^{-5}M_p$). In this situation, we have found that the values for the parameters are given by $A_1=5.08\times10^{14}M_p^{-3}$, $B_1=2.98\times10^{13}M_p^{-3}$ and $N_\text{end}=0.33$, respectively. In the lower left plot we have again used $\beta=1$ and for the brane $\sigma=10^{-7}M_p^4$ (or $\alpha_1=2.58\times10^{-4}M_p$), together with  the values $A_1=1.61\times10^{14}M_p^{-3}$, $B_1=3.56\times10^{13}M_p^{-3}$ and $N_\text{end}=0.31$. In the lower right plot, we have considered the standard Friedmann equation  in which the parameter $\beta=2$ i.e., $H^2\propto\rho$, and then $\alpha_2=1$. For this case, we have obtained that   the values for the parameters $A_1$, $B_1$ and $N_\text{end}$ become $A_1=2.07\times10^{10}M_p^{-2}$, $B_1=8.35\times10^{8}M_p^{-2}$ and $N_\text{end}=0.49$, respectively.

We mention that the stage of instantaneous reheating is given when the number of $e-$folds at the end of inflation $N_\text{reh}$ approaches  zero. In these panels, this corresponds to the point  in which all lines converge to the value
$N_\text{reh}\simeq 0$. From this plot, we note that the scenario of instantaneous reheating the model presents the maximum temperature of the reheating on the order of $T_\text{reh}\sim 10^{16}$GeV  for $\beta=2$ and for the standard Friedmann equation a little less $T_\text{reh}\sim 10^{15}$GeV. Besides, we observe that the stage of 
instantaneous reheating  (when $N_\text{reh}\simeq 0$) does not depend on the  EoS parameter $\omega_\text{reh}$, since all curves associated to the different $\omega_\text{reh}$ converge to the same point. Besides, we can note from the different panels that the highest value  of the
number of $e-$folds during the reheating takes place when the EoS parameter $\omega_\text{reh}=2/3$ independently of the value of $\beta$, and this value of $N_\text{reh}$ becomes $N_\text{reh}\lesssim 40$.

In relation to the reheating temperature  $T_\text{reh}$, we note that that the models associated to $\beta=1$ and $\beta=2$, present a good compatibility with Planck's 1$\sigma$ bounds on the index $n_s$, for the distinct values of the parameter $\omega_\text{reh}$, excluding the value $\omega_\text{reh}=-1/3$,  when the temperature $T_\text{reh}<10^{+12+13}$GeV. In particular,  when the EoS parameter takes the value $\omega_\text{reh}=-1/3$, the compatibility with Planck's 1$\sigma$ bounds is disadvantaged  when  
 the temperature 
$T_\text{reh}<10^{+13+14}$GeV for the case in which $\beta=2$,  
and when the temperature $T_\text{reh}<10^{+12+13}$GeV for the standard Friedmann equation ($\beta=1$).

The summary of the parameter-space  of the model $\mathcal{F}=\alpha_\lambda H^\beta$ can be consulted in the Table.\ref{Tabla1}. Here
the permitted values obtained for the integration constants $A_1$ and $B_1$ together with the number of $e-$folds at the end of inflation $N_\text{end}$  for  two different values of 
the parameter or power $\beta=1,2$
are shown in Table \ref{Tabla1}. In particular , for the situation in which $\beta=1$, we have considered three different values of the tension for the brane $\sigma$. Additionally, 
in this table we have utilized  the values $N_k=60$, $A_{s_k}=2.2\times 10^{-9}$ and $r_k=0.039$. Also, from this table we can  observe that the ratio between the integration constants $A_1$ and $B_1$ is the order of $\mathcal{O}(10^{2})$. Besides, we note from Table I that the number of $e-$folds at the end of the inflationary era $N_\text{end}\sim \mathcal{O}(1)$ independently on the value of the power $\beta$. 
Furthermore, we observe that the values associated to the integration constants $A_1$ and $B_1$ are very similar for the cases in which the  brane tension $\sigma= 10^{-8} M_p^4$ and $\sigma= 10^{-7} M_p^4$. Additionally, we note that when we decrease the value of the power $\beta$ from the values $\beta=2$ to $\beta=1$, the constraints on the integration constants $A_1$ and $B_1$ decrease by several orders of magnitude  $\sim \mathcal{O}(10^{4})$.

\begin{table}[H]
\centering
\begin{tabular}{|c|c|c|c|c|c|}
\hline
\,\,\,$\beta$\,\,\, & $\sigma \, \left[M_p^4\right]$ & $\alpha_\lambda \, \left[M_p^{2-\beta}\right]$  & $A_1\,\left[M_p^{\beta-4}\right]$ & $B_1\,\left[M_p^{\beta-4}\right]$ & $N_\text{end}$ \\ \hline
1 & $10^{-9}$ & $2.58\times10^{-5}$ & $1.61\times10^{15}$ & 
$1.15\times10^{13}$   & 0.33 \\ \hline
1 & $10^{-8}$ & $8.16\times10^{-5}$ & $5.08\times10^{14}$ & $2.98\times10^{13}$   & 0.33 \\ \hline
1 & $10^{-7}$ & $2.58\times10^{-4}$ & $1.61\times10^{14}$ & $3.56\times10^{13}$   & 0.31 \\ \hline
2 & - & 1                   & $2.07\times10^{10}$ & $8.35\times10^{8}$    & 0.49 \\ \hline
\end{tabular}
\caption{This table summarizes the parameter-space of the modified Friedmann equation $\mathcal{F}=\alpha_\lambda H^\beta$, for the specific cases in which the power $\beta$ takes the values $\beta=1$ and $\beta=2$, respectively. In particular for the situation in which $\beta=1$ we have considered three different values for the  brane tension $\sigma$.
} 
\label{Tabla1}
\end{table}

\subsection{Reconstruction example II: $\mathcal{F}(H)=H^2-\gamma H^{-2}$}\label{RecII}

As a  second example that we will consider to rebuild the background variables corresponds to the modified Friedmann equation in which the function $\mathcal{F}(H)$ motivated by  the  Kaniadakis Entropy \cite{Sheykhi:2023aqa}
is given by 
\begin{equation}
\label{F2}
\mathcal{F}(H)=H^2-\gamma H^{-2},   
\end{equation}
where $\gamma$ is a positive constant with dimensions of $M_p^4$.
As we comment in the Introduction, the 
Kaniadakis entropy (or K-entropy), is a one-parameter extension of the classical Boltzmann–Gibbs–Shannon entropy\cite{Sheykhi:2023aqa}. Besides,  we note that in order 
to obtain an energy density  $\rho>0$, we  have to satisfy that the Hubble parameter $H>\gamma^{1/4}$.  

For this type the function $\mathcal{F}(H)$, we find from Eq.(\ref{g(H)}) that the function $g(H)$ becomes
\begin{equation}
    g(H)=-\frac{7\gamma+3H^4}{\gamma+H^4}.
\end{equation}

Thus, introducing this result into Eq.(\ref{G(H)}) and using the attractor $n_s=n_s(N)$ given by Eq.(\ref{attractor}), we find that 
from  Eq.(\ref{eqdiff}),  the equation for the Hubble parameter in terms of the number of $e-$folds $N$ can be written as

\begin{equation}
    \frac{\gamma}{3}+H^4=2\left(\frac{A_2}{N}+B_2\right)H^6,
\end{equation}
where the quantities $A_2$ and $B_2$ are two integration constants. As we will see later the constant $A_2>0$ and for simplicity in the following we will consider that integration constant $B_2>0$.
By making a change of variable, such that $x=H^2$ in the previous equation, then we  obtain a cubic equation, whose real solution  for the Hubble parameter $H=H(N)$ is given by
\begin{equation}
\label{H(N)2}
H(N)=x^{1/2}= \left[\frac{1}{3 f}\left(1+\mathcal{A}^{1/3}+\mathcal{A}^{-1/3}\right)\right]^{1/2},
\end{equation}
where we have assumed that the function $\mathcal{A}=\mathcal{A}(N)$ in terms of the number of $e-$folds $N$ is a quantity positive and it is defined as 
\begin{equation}
\label{Adef}
   \mathcal{A}=\mathcal{A}(N)=\frac{1}{2}\left[2+9\gamma f^2+\sqrt{\left(2+9\gamma f^2\right)^2-4}\,\,\right]\,\,\,\mbox{where}\,\,\,f=f(N)=2\left(\frac{A_2}{N}+B_2\right).
\end{equation}

As before, in order to obtain the reconstruction of the background variables, we need to determine the relation between the number of $e-$folds and the scalar field, i.e., $N=N(\phi)$. In this way,  
we can  use Eq.(\ref{Nphi}) to find  the number of $e$-folds in terms of the scalar field $\phi$ from the differential equation given by 
\begin{equation}
    \frac{dN}{d\phi}=N_\phi=\sqrt{\frac{\kappa}{2A_2}}\left(\frac{N}{H}\right)=\sqrt{\frac{3\kappa}{2A_2}}\sqrt{2}N\left(1+\mathcal{A}^{1/3}+\mathcal{A}^{-1/3}\right)^{-1/2}\left(\frac{A_2}{N}+B_2\right)^{1/2}.\label{Nex}
\end{equation}
Here we note that this differential equation has no analytical solution. In this context and in order to solve the Eq.(\ref{Nex}), we can assume that during the inflationary scenario the ratio between the integration constants $A_2/B_2\ll N$. Thus, using this approximation, we obtain that the Hubble parameter as a function of the number of $e-$folds $N$ i.e., $H=H(N)$ can be approximate to the expression
\begin{equation}
    H(N)\simeq\mathcal{C}_1-\frac{\mathcal{C}_2}{N}+\mathcal{O}(N^{-2}),\label{HdeN2}
\end{equation}
where the constants $\mathcal{C}_1$ and $\mathcal{C}_2=(\bar{\mathcal{C}}_2+\bar{\mathcal{C}}_3)/\bar{\mathcal{C}}_4 $ are given   by
\begin{eqnarray}
\mathcal{C}_1&=&\left(1/\sqrt{6B_2}\right)\left[1+\left(1+2\mathcal{B}+2\sqrt{\mathcal{B} \left(\mathcal{B}+1\right)}\right)^{1/3}+\left(1+2\mathcal{B}+2\sqrt{\mathcal{B} \left(\mathcal{B}+1\right)}\right)^{-1/3}\right]^{1/2},\\
\bar{\mathcal{C}}_2&=&A_2\left[1+\left(1+2\mathcal{B}+2\sqrt{\mathcal{B} \left(\mathcal{B}+1\right)}\right)^{1/3}+\left(1+2\mathcal{B}+2\sqrt{\mathcal{B} \left(\mathcal{B}+1\right)}\right)^{-1/3}\right]\\
\bar{\mathcal{C}}_3&=&\frac{2A_2}{3}\frac{\left[\sqrt{\mathcal{B} \left(\mathcal{B}+1\right)}+2\mathcal{B}\left(1+\mathcal{B}+\sqrt{\mathcal{B}(\mathcal{B}+1)}\right) \right]\left[1-\left(1+2\mathcal{B}+2\sqrt{\mathcal{B}(\mathcal{B}+1)}\right)^{2/3}\right]}{(\mathcal{B}+1)\left(1+2\mathcal{B}+2\sqrt{\mathcal{B}(\mathcal{B}+1)}\right)^{4/3}}\\
\bar{\mathcal{C}}_4&=&2\sqrt{6}B_2^{3/2}\left[1+\left(1+2 \mathcal{B}+2 \sqrt{\mathcal{B} \left(\mathcal{B}+1\right)}\right)^{1/3}+\left(1+2\mathcal{B}+2 \sqrt{\mathcal{B} \left(\mathcal{B}+1\right)}\right)^{-1/3}\right]^{1/2} ,
\end{eqnarray}
with the quantity $\mathcal{B}$ defined as  $\mathcal{B}=9\gamma B_2^2$. 

Under the approximation in which the ratio  
$A_2/B_2\ll N$, we have assumed that the early universe suffers an inflationary expansion from a nearly  exponential expansion of the scale factor, in which the Hubble parameter is practically a constant and given by Eq.(\ref{HdeN2}),
 (for a quasi-de Sitter inflation, see e.g.,  Refs.\cite{Barrow:1990vx,Barrow:2007zr}).

Thus, considering  the approximation for the Hubble parameter given by   Eq.(\ref{HdeN2}), we find that the differential equation given by (\ref{Nex}) can be approximate to 
\begin{equation}
    N_\phi=\sqrt{\frac{\kappa}{2A_2}}\left(\frac{N}{H}\right)\simeq\sqrt{\frac{\kappa}{2A_2}}
\left(\frac{N}{\mathcal{C}_1-\mathcal{C}_2/N}\right).\label{eqr}
\end{equation}
In this form, from Eq.(\ref{eqr}), we obtain that 
the number of $e-$folds in terms of the scalar field $\phi$ i.e., the solution $N=N(\phi)$ can be written as
\begin{equation}
    N(\phi)\simeq -\frac{\mathcal{C}_2}{\mathcal{C}_1} \left(\frac{1}{\text{ProductLog}\left\{-\frac{\mathcal{C}_2}{\mathcal{C}_1}\exp{\left[-\frac{1}{\mathcal{C}_1}\left(\sqrt{\frac{\kappa}{2A_2}}\,\phi-\tilde{c}_2\right)\right]}\right\}}\right),
\end{equation}
where $\tilde{c}_2$ corresponds to a new integration constant. Besides, 
here the ProductLog function corresponds to a product logarithm, also
called the Omega function or Lambert W function,  see e.g., Ref.\cite{PLog}.

Thus, we find that the reconstruction of the Hubble parameter as a function of the scalar field becomes
\begin{equation}
H(\phi)\simeq \mathcal{C}_1\left(1+\text{ProductLog}\left\{-\frac{\mathcal{C}_2}{\mathcal{C}_1}\exp{\left[-\frac{1}{\mathcal{C}_1}\left(\sqrt{\frac{\kappa}{2A_2}}\,\phi-\tilde{c}_2\right)\right]}\right\}\right).
\end{equation}

Besides, the reconstruction of the effective potential in terms of the scalar field under the slow roll approximation  yields
\begin{equation}
V(\phi)\simeq \frac{3}{\kappa}\left[H^2(\phi)-\frac{\gamma}{H^2(\phi)}\right] \simeq \frac{3\mathcal{C}_1^2}{\kappa}\frac{
  \left[\left(1+\text{ProductLog}\left\{-\frac{\mathcal{C}_2}{\mathcal{C}_1}\exp{\left[-\frac{1}{\mathcal{C}_1}\left(\sqrt{\frac{\kappa}{2A_2}}\,\phi-\tilde{c}_2\right)\right]}\right\}\right)^4-\gamma/\mathcal{C}_1^4\right]}{\left(1+\text{ProductLog}\left\{-\frac{\mathcal{C}_2}{\mathcal{C}_1}\exp{\left[-\frac{1}{\mathcal{C}_1}\left(\sqrt{\frac{\kappa}{2A_2}}\,\phi-\tilde{c}_2\right)\right]}\right\}\right)^2}.   \label{Pot2}
\end{equation}

On the other hand, we can find a constraint on the integration constant $A_2$, using the expression for the power spectrum of the scalar perturbation. In this sense, 
we have that the amplitude of the power spectrum of the scalar perturbation, given by Eq.(\ref{AsdeH}) becomes
\begin{equation}
\label{AsModel2}
    A_s=\frac{\kappa}{8\pi^2}\frac{N^2}{A_2}\,\,\,\Rightarrow \,\,\,A_2=\frac{\kappa}{8\pi^2}\frac{N^2}{A_s}.
\end{equation}
Thus, from  Eq.(\ref{AsModel2}) and assuming that the number of $e-$folds $N_k=60$ and the power spectrum of the scalar perturbation corresponds to $A_s=2.2\times 10^{-9}$ we find that the integration constant 
$A_2=2.07\times 10^{10}M_p^{-2}$. 

In addition, using Eq.(\ref{HdeN2}), we find that the tensor to scalar ratio in terms of the number of $e-$folds $N$ yields
\begin{equation}
    r=16A_2\frac{H^2}{N^2}\simeq \frac{2\kappa}{\pi^2A_s}\left(\mathcal{C}_1-\frac{\mathcal{C}_2}{N}\right)^2.\label{rN2}
\end{equation}
Here we note that using Eq.(\ref{rN2}),  we will numerically obtain a constraint on the second integration constant $B_2$, since 
 the ratio $r$  is evaluated at the horizon exit during the inflationary epoch where it  is constrained by $\left.r(k)\right|_{k=a_kH_k}=r_k<0.039$ together with the number $N_k=60$. Thus, numerically we will determine the value of the integration constant $B_2$.

Besides, by considering that the function $\mathcal{F}(H)=H^2-\gamma H^{-2}>0$ or $H^4>\gamma$, then  combining Eqs.(\ref{AsModel2}) and (\ref{rN2}) we obtain an upper bound for the  parameter $\gamma$ given by 
\begin{equation}
\left(\frac{r\,N^2}{16\, A_2}\right)^2>\gamma.
\end{equation}
In particular considering the upper limit for the tensor to scalar ratio  $r_k=0.039$ together with the number of $e-$folds $N_k=60$ and the constraint found  on the integration constant  $A_2=2.07\times10^{10}M_p^{-2}$ we have 
\begin{equation}
1.80\times 10^{-19} \,M_{p}^4>\gamma.\label{eq1}
\end{equation}

\begin{figure}[h]
    \centering
\includegraphics[width=0.9\linewidth]{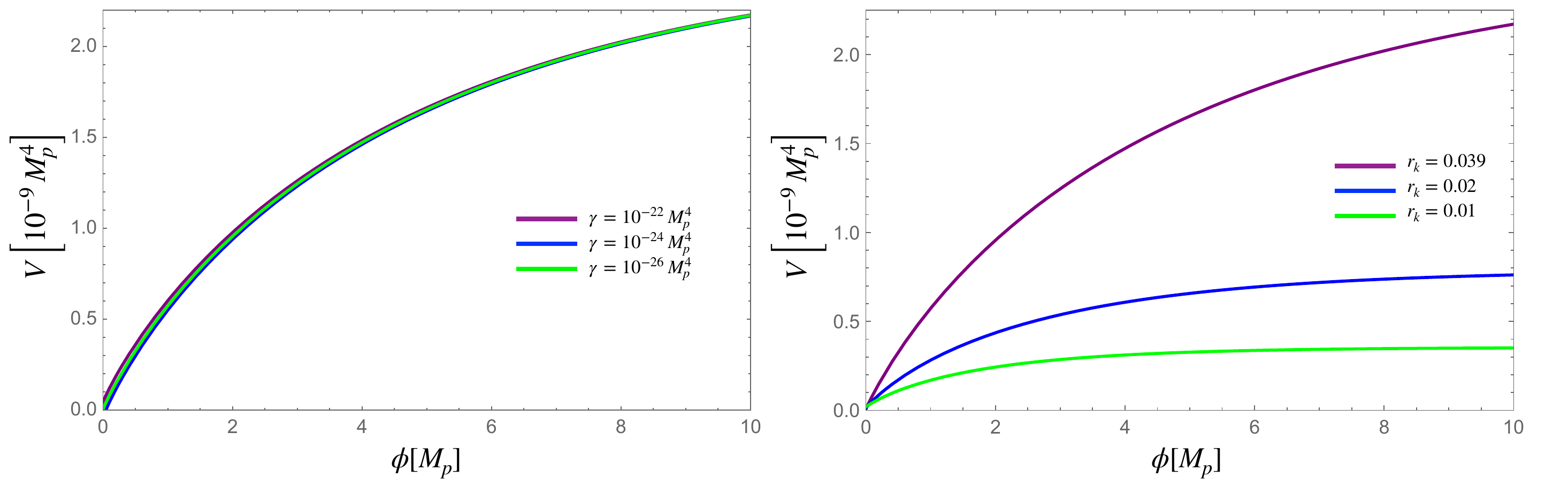}
    \caption{The left panel shows the reconstructed effective potential in terms of the scalar field for three different values of the parameter $\gamma$, when we  fix the tensor to scalar ratio to the upper bound $r_k=0.039$. The right panel also shows the reconstructed effective potential as a function  of the scalar field for three different values of the tensor to scalar ratio $r$, when we  now set the parameter $\gamma=10^{-24}M_p^4$. In both panels we have used that the constant $\tilde{c}_2=0$ for simplicity.
    } 
    \label{Fig_Potential_M2}
\end{figure}

In Fig.\ref{Fig_Potential_M2}, the left panel shows the reconstructed effective potential in terms of the scalar field given by Eq.(\ref{Pot2}) for different values of the parameter $\gamma$, when we  fix the tensor to scalar ratio $r_k=0.039$. From this panel, we note that  for the different values of the parameter $\gamma$, which satisfy the upper bound given by Eq.(\ref{eq1}) in which $10^{-19}M_p^4>\gamma$, the shape of the reconstructed effective potential $V(\phi)$ does not change, since  the different lines associated to  $\gamma$ are overlaid one after the other. This suggests that  due to that the parameter $\gamma\ll 1$, the reconstruction of the effective potential $V(\phi)$  is not strongly affected for this parameter. To build this panel,  we have used three different values of the parameter $\gamma$ where we fix the value of the tensor to scalar ratio $r_k=0.039$. The purple curve corresponds to  $\gamma= 10^{-22}M_p^4$, and the values obtained of $B_2$, $\mathcal{C}_1$ and $\mathcal{C}_2$ result $B_2=5.95\times10^8M_p^{-2}$, $\mathcal{C}_1=2.90\times10^{-5}M_p$ and $\mathcal{C}_2=5.05\times10^{-4}M_p$, respectively. To obtain the blue curve we have used $\gamma= 10^{-24}M_p^4$, and then $B_2=5.94\times10^8M_p^{-2}$, $\mathcal{C}_1=2.90\times10^{-5}M_p$ and $\mathcal{C}_2=5.06\times10^{-4}M_p$. Finally, the green curve is obtained using $\gamma= 10^{-26}M_p^4$, $B_2=5.94\times10^8M_p^{-2}$, $\mathcal{C}_1=2.90\times10^{-5}M_p$ and $\mathcal{C}_2=5.06\times10^{-4}M_p$. It is useful to remember that the value of the parameter $B_2$ is obtained numerically from Eq.(\ref{rN2}), fixing the tensor to scalar ratio $r$ for the different values of $\gamma$. In particular for the tensor to scalar ratio $r_k=0.039$, we find that of the integration constant $B_2$ does not change significantly for the different values of $\gamma$ utilized.


In the right panel of the Fig.\ref{Fig_Potential_M2}, we also show the reconstructed effective potential 
$V$ as a function of the scalar field $\phi$. Here three curves have been drawn for different values of the tensor to scalar ratio at the time of the crossing, when  we fix  the parameter $\gamma$ to the value $\gamma=10^{-24}M_p^4$. In particular, the purple curve corresponds to the value $r_{k}=0.039$, and then  we numerically find that the integration constant $B_2$ becomes
 $B_2=5.94\times10^{8}M_p^{-2}$ and therefore the quantities  $\mathcal{C}_1$ and $\mathcal{C}_2$ result $\mathcal{C}_1=2.90\times10^{-5}M_p$ and $\mathcal{C}_2=5.06\times10^{-4}M_p$, respectively.
In the case of the  blue curve in which $r_{k}=0.02$, we find that the parameters are $B_2=1.90\times10^{9}M_p^{-2}$, $\mathcal{C}_1=1.62\times10^{-5}M_p$ and $\mathcal{C}_2=8.85\times10^{-5}M_p$. Finally, the green curve corresponds to $r_{k}=0.01$, and then the values are given by $B_2=4.24\times10^{9}M_p^{-2}$, $\mathcal{C}_1=1.09\times10^{-5}M_p$ and $\mathcal{C}_2=2.65\times10^{-5}M_p$, respectively.

On the other hand,   we will analyze the reheating era for our second example, in which the function $\mathcal{F}(H)$ associated to the modified Friedmann equation is given by $\mathcal{F}(H)=H^2-\gamma H^{-2}$. In the same way as in the example I, we  solve the Eq.(\ref{EqRhoEnd}) using this function  $\mathcal{F}(H)$ to find that the energy density at the end of the inflationary epoch becomes

\begin{equation}
\label{rhoend2}
    \rho_\text{end}=\frac{3}{8}V_\text{end}\left(3\pm\sqrt{1+\frac{32\gamma}{\kappa^2V_\text{end}^2}}\,\,\right).
\end{equation}
In order to obtain the standard result of the above expression, in which $\rho_\text{end}\rightarrow (3/2)V_\text{end}$, when $\gamma\rightarrow 0$, it is necessary to consider the positive sign of the solution for $\rho_\text{end}$ given by Eq.(\ref{rhoend2}). Thus, in the following we will consider the positive sign of the solution for $\rho_\text{end}$ of  Eq.(\ref{rhoend2}).

\begin{figure}[h]
    \centering
    \includegraphics[width=0.45\linewidth]{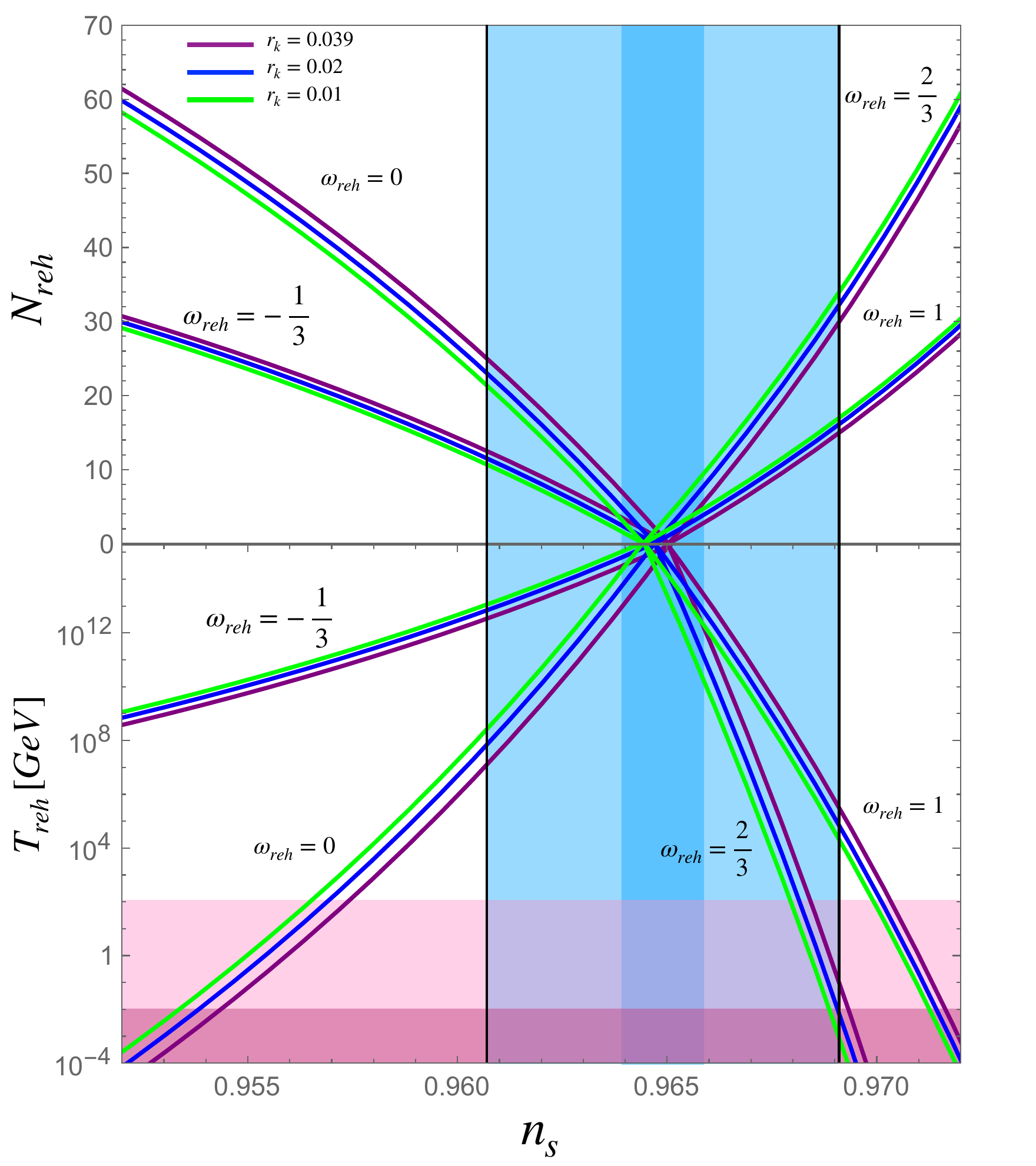}
    \caption{The upper panel shows the number of $e-$folds and the lower panel shows the temperature during the reheating stage versus the scalar spectral index, for different values of the tensor to scalar ratio. In these panels the different curves and shaded regions are similar to  the Fig.(\ref{Fig_Reh_M1}). Besides,  we have fixed the parameter $\gamma=10^{-24}M_p^4$.} 
    \label{Fig_Reh_M2}
\end{figure}

Besides, as before  using  Eqs.(\ref{Treh}), (\ref{Nreh}) and (\ref{attractor}) and considering  Eqs.(\ref{rN2}) and (\ref{rhoend2}), we can rewrite the reheating temperature $T_\text{reh}$ and the number of $e-$folds  $N_\text{reh}$ during the  reheating epoch as a function of the spectral index $n_s$. In this way, we find that $T_\text{reh}$ and $N_\text{reh}$ in terms of the scalar spectral index can be written as
\begin{equation}
\label{TrehM2}
	T_\text{reh}(n_s)=\text{exp}\left[-\frac{3}{4}(1+\omega_\text{reh})N_\text{reh}(n_s)\right]\left[\frac{45\,V_\text{end}\left(3+\sqrt{1+\frac{32\gamma}{\kappa^2V_\text{end}^2}}\,\,\right)}{4g_{\star\text{,reh}}\pi^2}\right]^{1/4},\,\,\,\mbox{and}
\end{equation}

\begin{equation}
\label{NrehM2}
N_\text{reh} =\frac{4}{1-3\omega_\text{reh}}\left[\frac{2}{n_s-1}-\ln\left(\frac{k}{a_0T_0}\right)-\frac{
\ln}{3}\left(\frac{11g_\text{s,reh}}{43}\right)
	-\frac{\ln}{4}\left(\frac{45\kappa^2\,V_\text{end}\left(3+\sqrt{1+\frac{32\gamma}{\kappa^2V_\text{end}^2}}\,\right)}{4g_{\star\text{,reh}}\pi^2}\right)+\tilde{r}_2(n_s)\right], 
\end{equation}
where $\omega_\text{reh}\neq 1/3$  and the function $\tilde{r}_2(n_s)$ is defined as
 \begin{equation}
    \tilde{r}_2(n_s)=\ln\left\{\kappa^{1/2}\left[\mathcal{C}_1-\frac{(1-n_s)}{2}\mathcal{C}_2\right]\right\}.
\end{equation}
The Fig.\ref{Fig_Reh_M2} shows the number of $e-$folds  (upper panel) and the 
temperature (lower panel) during the reheating scenario as functions of the spectral index for the modified Friedmann equation given by Eq.(\ref{F2}). In these panels we have used different values for the tensor to scalar ratio as well different values associated to the EoS parameters $\omega_\text{reh}$ in each case. Thus, 
 three curves have been plotted, one for each value of $r$, for each EoS parameter  $\omega_\text{reh}$ with values  $\omega_\text{reh}=\{-1/3,0,1,2/3\}$. As before, we note that the instantaneous reheating of the model 
 takes place when $N_\text{reh}\sim 0$ and temperature corresponds to a maximum valor. Besides, we note that any temperature between the BBN bound and the  instantaneous reheating value is allowed inside the Planck's 1$\sigma$ bound, 
independently of the EoS parameter. However, we note that  the curve associated with the upper limit for the tensor to scalar ratio $r_k\sim 0.039$ (purple curve) is more centered around the value within  the Planck's 1$\sigma$. Besides, we observe that the model predicts from Planck data at 1$\sigma$ limits ($n_s=0.9649\pm 0.042$) a small number of $e-$folds during the reheating epoch $N_\text{reh}<35$ for the different temperatures $T_\text{reh}$.

\begin{figure}[h]
    \centering
    \includegraphics[width=0.65\linewidth]{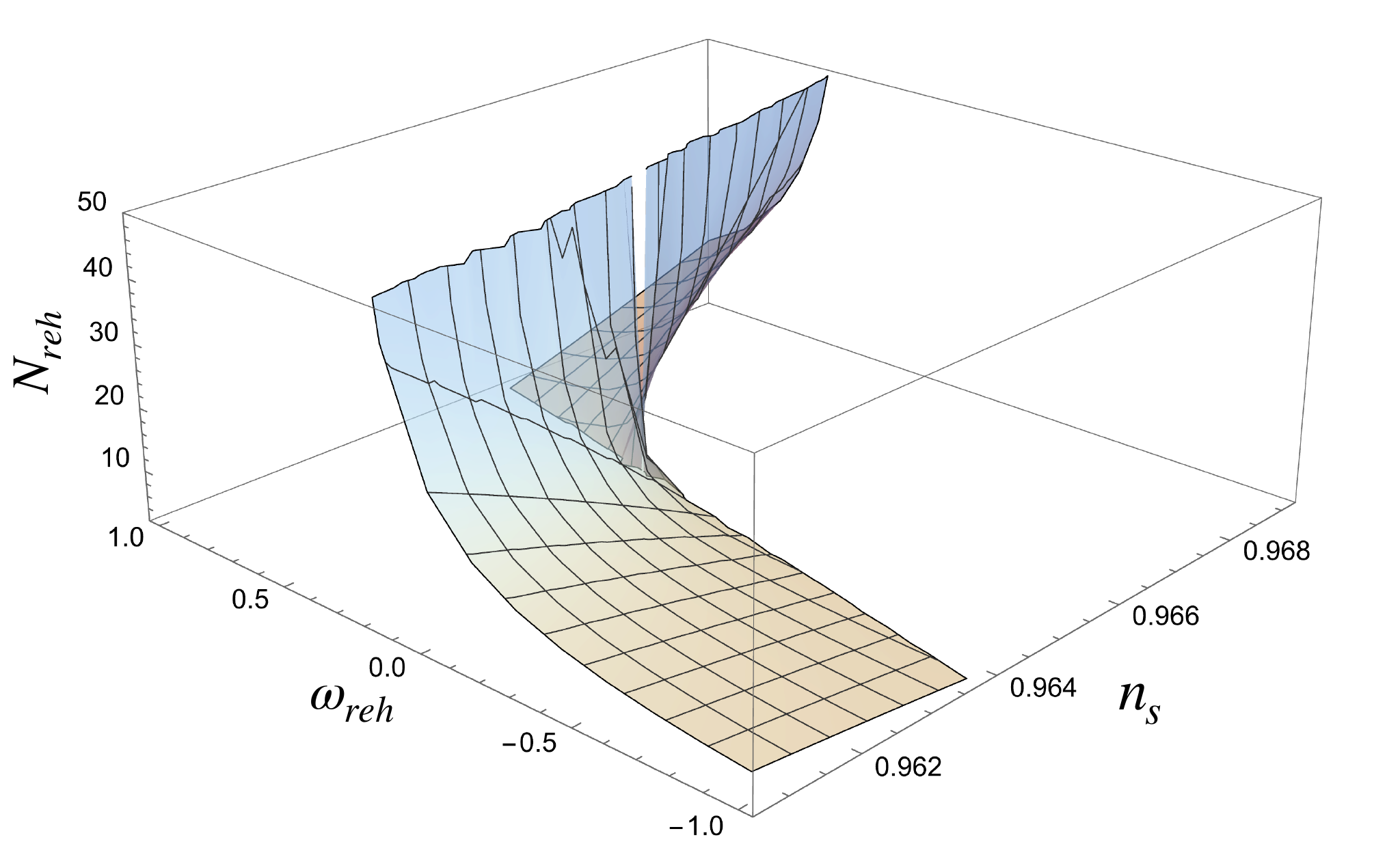}
    \caption{ Three-dimensional plot for the number of $e-$folds versus  the EoS parameter $\omega_\text{reh}$ and the scalar spectral index $n_s$.
    Here we have used that the tensor to scalar ratio corresponds to the upper bound $r_k=0.039$ and $N_k=$60.} 
    \label{Fig_Plot3DNreh_M2}
\end{figure}

The Fig.\ref{Fig_Plot3DNreh_M2} shows the three-dimensional plot for the number of $e-$folds in terms of the EoS parameter and the scalar spectral index, given by Eq.(\ref{NrehM2}). From this figure, we note that the maximum values of the number of $e-$folds occur for values of the EoS parameter in the range  $0<\omega_\text{reh}<1$ within the Planck's 1$\sigma$ bound. Besides, from this 3-D plot we observe that the number of the $e-$folds, during the reheating epoch becomes  smaller, when we consider negative values of the EoS parameter $\omega_\text{reh}$.

\subsection{Reconstruction example III: $\mathcal{F}(H)=H^2\pm \theta H^{4}$ }\label{RecIII}

As a final example to reconstruct the inflationary scenario using the scalar spectral index $n_s=n_s(N)$,  
 we will analyze  the modified Friedmann equation given by 
 \begin{equation}
    \mathcal{F}(H)=H^2\pm\theta H^4,\label{59}
\end{equation}
where $\theta$ is a positive constant with dimension $M_p^{-2}$. 
This type of modification to the Friedmann equation comes for example, from the Chern-Simons theories of gravity \cite{Gomez:2021opp} or quantum corrections to the entropy of the apparent horizon \cite{Cai:2008ys}  as well  black hole entropy in loop quantum gravity\cite{Ashtekar:1997yu,Kaul:2000kf}  or due to  correction to black hole entropy due to thermal
equilibrium fluctuation or quantum fluctuation, see also Refs.\cite{Solodukhin:1997yy,Kastrup:1997iu}.  

In relation to the Eq.(\ref{59}) and the parameter $\theta$, we need to ensure that   the term 
$H^2$ should be comparable to the term 
$\lvert\theta\lvert H^4$.  In this form, we can assume that  the parameter $\theta$ must be of the order  $\lvert\theta\lvert\sim H^{-2}$ for both terms to be significant during the early universe.

To begin the reconstruction of the background variables from the modified Friedmann equation defined by Eq.(\ref{59}), we determine 
 the expression of the function $g(H)$ given by Eq.(\ref{g(H)}) yields
\begin{equation}
    g(H)=-\frac{(3\pm 2\theta H^2)}{\left(1\pm 2\theta H^2\right)}.
\end{equation}
Now using Eqs.(\ref{eqdiff}),  (\ref{G(H)}) and (\ref{attractor}), we find that the first order differential equation for the Hubble parameter as a function of the number of $e-$folds $N$ can be written as 
 
\begin{equation}
    \left(1\pm2\theta H^2\right)\frac{H_N}{H^3}=\frac{A_3}{N^2},\label{62}
\end{equation}
where $A_3$ is a  constant of integration and as we will see later this constant is positive. 
In the following, by simplicity,  we will consider the positive sign of Eq.(\ref{59}) in our analysis.

From Eq.(\ref{62})  we  obtain that the  solution for the Hubble parameter as a function of $N$  becomes
\begin{equation}
\label{H(N)III}
    H(N)=\frac{1}{\sqrt{2\theta}}\left[\text{ProductLog}\left(\frac{1}{2\theta}e^{\frac{1}{\theta}\left(\frac{A_3}{N}+B_3\right)} \right)\right]^{-1/2},
\end{equation}
where $B_3$ corresponds to a new integration constant. In the following,  we will assume that this constant $B_3>0$ for simplicity.

To rebuild the background variables, we need 
to find the number of $e-$folds $N$ as a function of $\phi$. Thus, from  Eq.(\ref{Nphi}), we have that the first order  differential equation for $N=N(\phi)$ becomes
\begin{equation}
    N_\phi=\sqrt{\frac{\kappa}{2A_3}}\left(\frac{N}{H}\right)=\sqrt{\frac{\kappa\,\theta}{A_3}}\,N\,\left[\text{ProductLog}\left(\frac{1}{2\theta}e^{\frac{1}{\theta}\left(\frac{A_3}{N}+B_3\right)} \right)\right]^{1/2}.\label{Np}
\end{equation}
Nevertheless, we cannot analytically solve this first order differential equation to obtain $N=N(\phi)$. In order to find an analytical solution to Eq.(\ref{Np}), we can consider that during inflation the ratio between the integration constants $A_3/B_3\ll N$. By using this approximation, we obtain that the Hubble parameter in terms of the number of $e-$folds can be approximate to an expansion quasi de Sitter given by 
\begin{equation}
     H(N)\simeq \mathcal{D}_1-\frac{\mathcal{D}_2}{N}+\mathcal{O}(N^{-2}),\label{H3}
\end{equation}
where $\mathcal{D}_1$ and $\mathcal{D}_2$ correspond to two constants defined as 
\begin{eqnarray}
    \mathcal{D}_1&=&\frac{1}{\sqrt{2\theta}}\left[\text{ProductLog}\left(\frac{1}{2\theta}e^{B_3/\theta}\right)\right]^{-1/2},\,\,\,\,\,\,\,\mbox{and}\\
    \mathcal{D}_2&=&\frac{A_3}{2\sqrt{2}\theta^{3/2}}\left[\text{ProductLog}\left(\frac{1}{2\theta}e^{B_3/\theta}\right)\right]^{-1/2}\left[1+\text{ProductLog}\left(\frac{1}{2\theta}e^{B_3/\theta}\right)\right]^{-1},
\end{eqnarray}
respectively.
In this context, the  differential equation given by Eq.(\ref{Np}) is reduced to 

\begin{equation}
        N_\phi\simeq \sqrt{\frac{\kappa}{2A_3}}\,\left(\frac{N}{\mathcal{D}_1-\mathcal{D}_2/N}\right),
\end{equation}
and we find that the  solution  for the number of $e-$folds  $N$ as a function of the scalar field $\phi$ yields
\begin{equation}
    N(\phi)\simeq -\frac{\mathcal{D}_2}{\mathcal{D}_1} \left(\text{ProductLog}\left\{-\frac{\mathcal{D}_2}{\mathcal{D}_1}\exp{\left[-\frac{1}{\mathcal{D}_1}\left(\sqrt{\frac{\kappa}{2A_3}}\,\phi-\tilde{c}_3\right)\right]}\right\}\right)^{-1},\label{Np1}
\end{equation}
where $\tilde{c}_3$ is a new integration constant. In this form, replacing Eq.(\ref{Np1}) into Eq.(\ref{H3}),
we obtain that the reconstruction of the Hubble parameter 
$H=H(\phi)$, during the inflationary epoch results
\begin{equation}
H(\phi)\simeq \mathcal{D}_1\left(1+\text{ProductLog}\left\{-\frac{\mathcal{D}_2}{\mathcal{D}_1}\exp{\left[-\frac{1}{\mathcal{D}_1}\left(\sqrt{\frac{\kappa}{2A_3}}\,\phi-\tilde{c}_3\right)\right]}\right\}\right).
\end{equation}

Besides, the reconstructed effective potential in terms of the scalar field can be written as 
\begin{eqnarray}
V(\phi)&\simeq&\frac{\left(3\mathcal{D}_1^2/\kappa\right)\left(1+\text{ProductLog}\left\{-\frac{\mathcal{D}_2}{\mathcal{D}_1}\exp{\left[-\frac{1}{\mathcal{D}_1}\left(\sqrt{\frac{\kappa}{2A_3}}\,\phi-\tilde{c}_3\right)\right]}\right\}\right)^2}{\left[1-\theta\,\mathcal{D}_1^2\left(1+\text{ProductLog}\left\{-\frac{\mathcal{D}_2}{\mathcal{D}_1}\exp{\left[-\frac{1}{\mathcal{D}_1}\left(\sqrt{\frac{\kappa}{2A_3}}\,\phi-\tilde{c}_3\right)\right]}\right\}\right)^2\right]^{-1}}.
\end{eqnarray}

Additionally, from the power spectrum of the scalar perturbations defined by Eq.(\ref{As}), we can determine a constraint on the parameter $A_3$ given by 
\begin{equation}
    A_s=\frac{\kappa}{8\pi^2}\frac{N^2}{A_3}\,\,\,\,\Longrightarrow\,\,\,\,A_3=\frac{\kappa}{8\pi^2}\frac{N^2}{A_s}>0.
\end{equation}    
Also, we find that the tensor to scalar ratio $r$ in terms of the number of $e-$folds $N $ becomes    
\begin{equation}
    r=16A_3\frac{H^2}{N^2}\simeq\frac{2\kappa}{\pi^2A_s}\left(\mathcal{D}_1-\frac{\mathcal{D}_2}{N}\right)^2\label{r3N},
\end{equation}

From Eq.(\ref{r3N}), different values of the parameter $B_3$ can be 
numerically obtain for various   values of the parameter $\theta$.
 
\begin{figure}[h]
    \centering
    \includegraphics[width=0.8\linewidth]{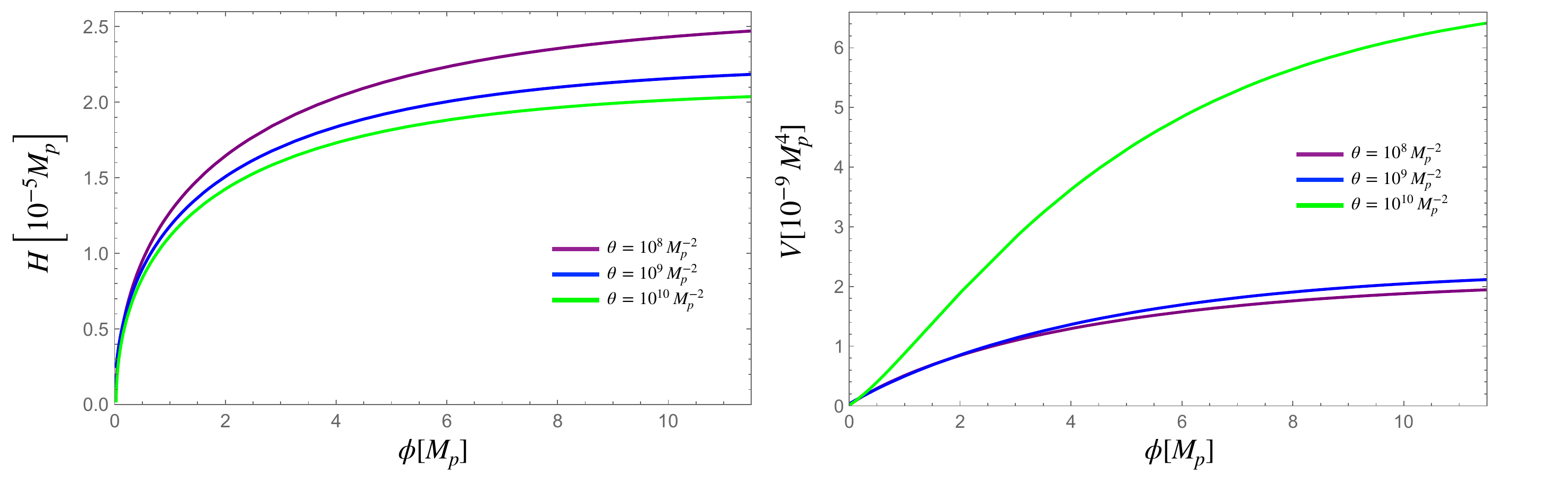}
    \caption{The figure shows the reconstructions of the Hubble parameter (left panel) and the effective potential (right panel) versus the scalar field, for different values of the parameter $\theta$. In both panels we have considered that the tensor to scalar ratio $r_k=0.039$ and the number of $e-$folds $N_k=$60. } 
    \label{Fig_Potential_M3}
\end{figure}

In Fig.\ref{Fig_Potential_M3} the left panel shows the reconstruction of the Hubble parameter versus the scalar field, for different values of the parameter $\theta$. In order to chose the different values of the parameter $\theta$, we have considered that $\theta\sim H^{-2}$ to ensure that the terms $H^2$ and $\theta H^4$ are the order. Besides, in the right panel we show the reconstruction of the effective potential  in terms of the inflaton field, for various values of the parameter $\theta$.
In the two panels we have fixed the tensor to scalar ratio $r_k=0.039$. Also, 
in both panels the purple curve is obtained using the following parameters; for $\theta=10^8M_p^{-2}$,  we determine that the constants $B_3$, $\mathcal{D}_1$  and $\mathcal{D}_2$ are given by $B_3=2.86\times10^{9}M_p^{-2}$, $\mathcal{D}_1=2.58\times10^{-5}M_p$ and $\mathcal{D}_2=3.15\times10^{-4}M_p$. The blue curve is obtained for the special value of  $\theta=10^9M_p^{-2}$, in which  $B_3=2.24\times10^{10}M_p^{-2}$, $\mathcal{D}_1=2.25\times10^{-5}M_p$ and $\mathcal{D}_2=1.18\times10^{-4}M_p$. Finally, the green curve is built    with the values;  $\theta=10^{10}M_p^{-2}$, $B_3=2.17\times10^{11}M_p^{-2}$, $\mathcal{D}_1=2.09\times10^{-5}M_p$ and $\mathcal{D}_2=1.94\times10^{-5}M_p$, respectively. In relation to the left panel of this figure, we note that the reconstructed Hubble parameter in terms of the scalar field presents an approximately   quasi de Sitter behavior (flat region)
 for large-$\phi$ ($\phi>10 M_p$). In addition,   
from  the right panel, we note that the reconstructed effective potential in terms of the scalar field $V=V(\phi)$ shows a maximum value given by a flat region for values of $\phi>10 M_p$, in which the scalar field begins to roll towards values of the scalar field $\phi\sim $0 in which  the inflationary epoch ends.

On the other hand, to analyze the reheating epoch in our modified Friedmann equation 
 $\mathcal{F}(H)=H^2+\theta H^{4}$, we will proceed as in the previous examples.  To determine the energy density at the end of the inflationary scenario, we need to calculate the second term of the Eq.(\ref{EqRhoEnd}). In this way, we have that this term becomes
\begin{equation}
 \frac{1}{2\kappa}\left.\left(\frac{d\mathcal{F}}{d\ln H}\right)\right|_\text{end}=\frac{1}{\kappa}H^2\left.\left(1+2\theta H^2\right)\right|_\text{end}.  
\end{equation}

Introducing this expression into the Eq.(\ref{EqRhoEnd}), we find that the energy density at the end of the inflationary stage results
\begin{equation}
\label{rhoend3}
    \rho_\text{end}=\frac{3}{\kappa\theta}\left(1+\kappa\theta V_\text{end}\pm\sqrt{1+\kappa\theta V_\text{end}}\,\,\right).
\end{equation}
Here we note that in order to obtain the standard result from Eq.(\ref{rhoend3}) in which $\rho_\text{end}=(3/2)V_\text{end}$, when $\theta\rightarrow 0$ , it is necessary to consider the negative sign of this equation. In the following, we will use the negative sign of the expression given by Eq.(\ref{rhoend3}).

 To study the reheating epoch, as before, we  need to determine the functional form of the reheating parameters $T_\text{reh}$ and $N_\text{reh}$ in terms of the scalar spectral index $n_s$  for our $\mathcal{F}(H)$-model. In this context, using  Eqs.(\ref{Treh}), 
  (\ref{Nreh}) and (\ref{attractor})    and now replacing  the new expressions given by 
  (\ref{r3N}) and (\ref{rhoend3}),
  we find that the temperature and the number of $e-$folds as a function of the $n_s$ during the reheating epoch result
\begin{equation}
\label{TrehM2}
	T_\text{reh}(n_s)=\text{exp}\left[-\frac{3}{4}(1+\omega_\text{reh})N_\text{reh}(n_s)\right]\left[\frac{90\left(1+\kappa\theta V_\text{end}-\sqrt{1+\kappa\theta V_\text{end}}\,\,\right)}{\kappa\theta\pi^2 g_{\star\text{,reh}} }\right]^{1/4},\,\,\,\mbox{and}
\end{equation}

\begin{equation}
\label{NrehM3}
N_\text{reh}=\frac{4}{1-3\omega_\text{reh}}\left[\frac{2}{n_s-1}-\ln\left(\frac{k}{a_0T_0}\right)-\frac{1}{3}\ln\left(\frac{11g_\text{s,reh}}{43}\right)
	-\frac{1}{4}\ln\left(\frac{90\kappa\left(1+\kappa\theta V_\text{end}-\sqrt{1+\kappa\theta V_\text{end}}\,\right)}{\theta\pi^2 g_{\star\text{,reh}} }\right)+\tilde{r}_3\right], 
\end{equation}
with the EoS parameter $\omega_\text{reh}\neq1/3$ and  the function $\tilde{r}_3=\tilde{r}_3(n_s)$   defined as
 \begin{equation}
    \tilde{r}_3(n_s)=\ln\left\{\kappa^{1/2}\left[\mathcal{D}_1-\frac{(1-n_s)}{2}\mathcal{D}_2\right]\right\}.
\end{equation}

\begin{figure}[h]
    \centering
\includegraphics[width=0.6\linewidth]{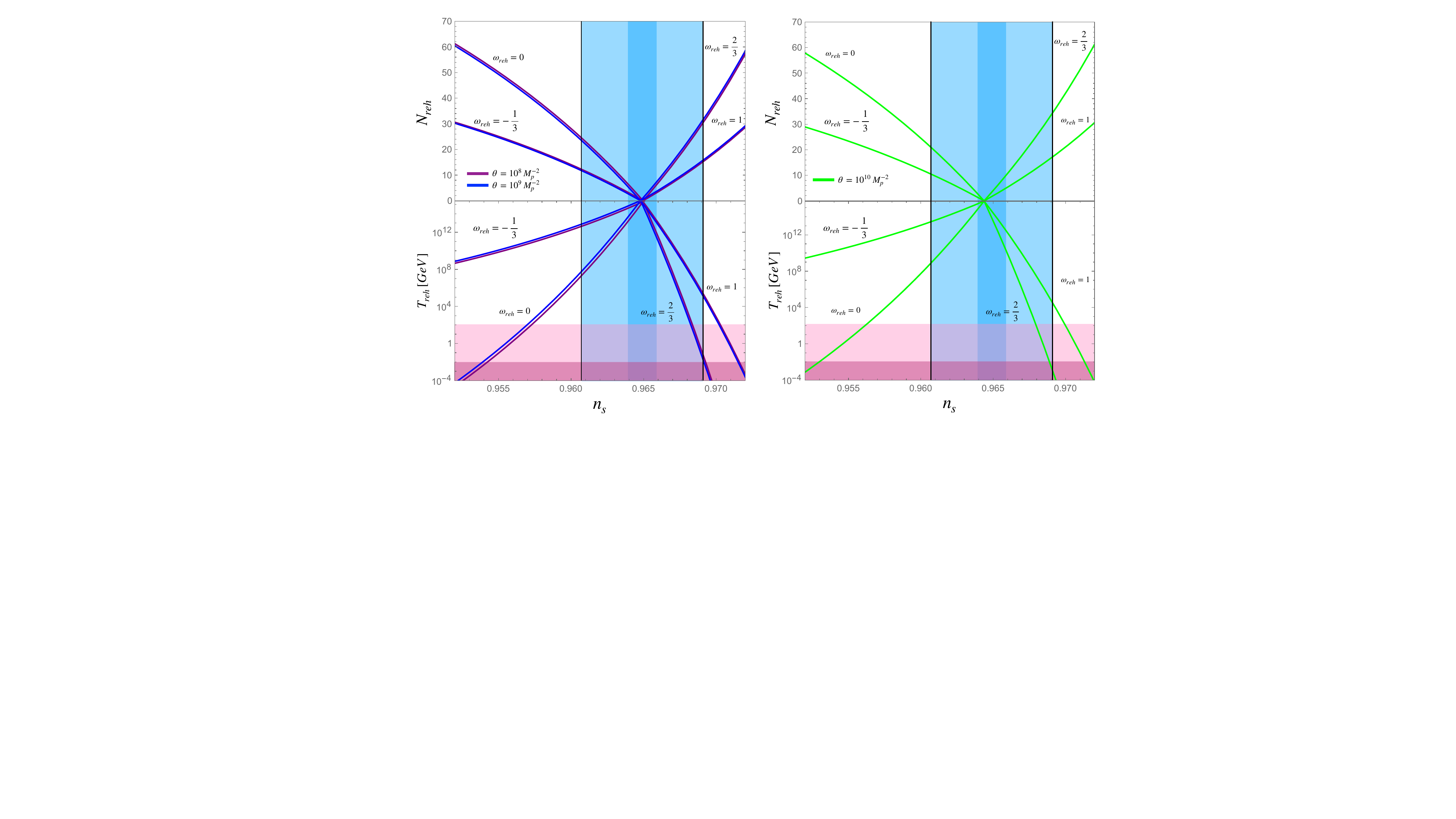}
    \caption{In both plots; the upper panel  shows the number of $e-$folds and the lower panel exhibits  the temperature during the reheating stage versus the scalar spectral index, for different values of the parameter $\theta$, as well as various values of the EoS parameter $\omega_\text{reh}$. In the left panel, we show two different values of $\theta$; $\theta=10^{8}M_p^{-2}$ and $\theta=10^{9}M_p^{-2}$, while the right panel shows the situation in which $\theta=10^{10}M_p^{-2}$.
    In these panels the different curves and shaded regions are as  the Fig.(\ref{Fig_Reh_M1}). Beside, in these plots,  we have fixed the tensor to scalar ratio $r_k=0.039$ together with $N_k=$60.
    } 
    \label{Fig_Reh_M3}
\end{figure}

The Fig.\ref{Fig_Reh_M3} exhibits  the number of $e-$folds  (upper panel) and the 
temperature (lower panel) during the reheating epoch versus the spectral index for the modified Friedmann equation $\mathcal{F}(H)=H^2+\theta H^4$. In these plots we have considered various values for the parameter $\theta$, as well various values associated to the EoS parameters $\omega_\text{reh}$ in each case. In the left panel, we have assumed two different values of the parameter $\theta$ for each EoS parameter  $\omega_\text{reh}$ with values of  $\omega_\text{reh}=\{-1/3,0,1,2/3\}$. In the right panel we have utilized the specific case in which the parameter $\theta= 10^{10}M_p^{-2}$ , also for  various values of the EoS parameter $\omega_\text{reh}$.

In this context, for each barotropic index $\omega_\text{reh}=\{-1/3,0,1,2/3\}$ three sets of curves have been drawn considering  different values of $\theta$.
In the left panel, the purple curve has been obtained by using the following set of parameters; for   $\theta=10^8M_p^{-2}$,  we have  found that the constants $B_3$, $\mathcal{D}_1$ and $\mathcal{D}_2$ together with the number of $e-$folds at the end of the inflationary era are given by
$B_3=2.86\times10^{9}M_p^{-2}$, $\mathcal{D}_1=2.58\times10^{-5}M_p$, $\mathcal{D}_2=3.15\times10^{-4}M_p$ and $N_\text{end}=13.1$, respectively. The blue curve corresponds to the value $\theta=10^9M_p^{-2}$, in which  $B_3=2.24\times10^{10}M_p^{-2}$, $\mathcal{D}_1=2.25\times10^{-5}M_p$, $\mathcal{D}_2=1.18\times10^{-4}M_p$ and $N_\text{end}=6.08$. In the right panel, 
 the green curve  corresponds to the values of $\theta=10^{10}M_p^{-2}$, $B_3=2.17\times10^{11}M_p^{-2}$, $\mathcal{D}_1=2.09\times10^{-5}M_p$,  $\mathcal{D}_2=1.94\times10^{-5}M_p$ and $N_\text{end}=1.53$, respectively.

\begin{figure}[h]
    \centering
    \includegraphics[width=0.5\linewidth]{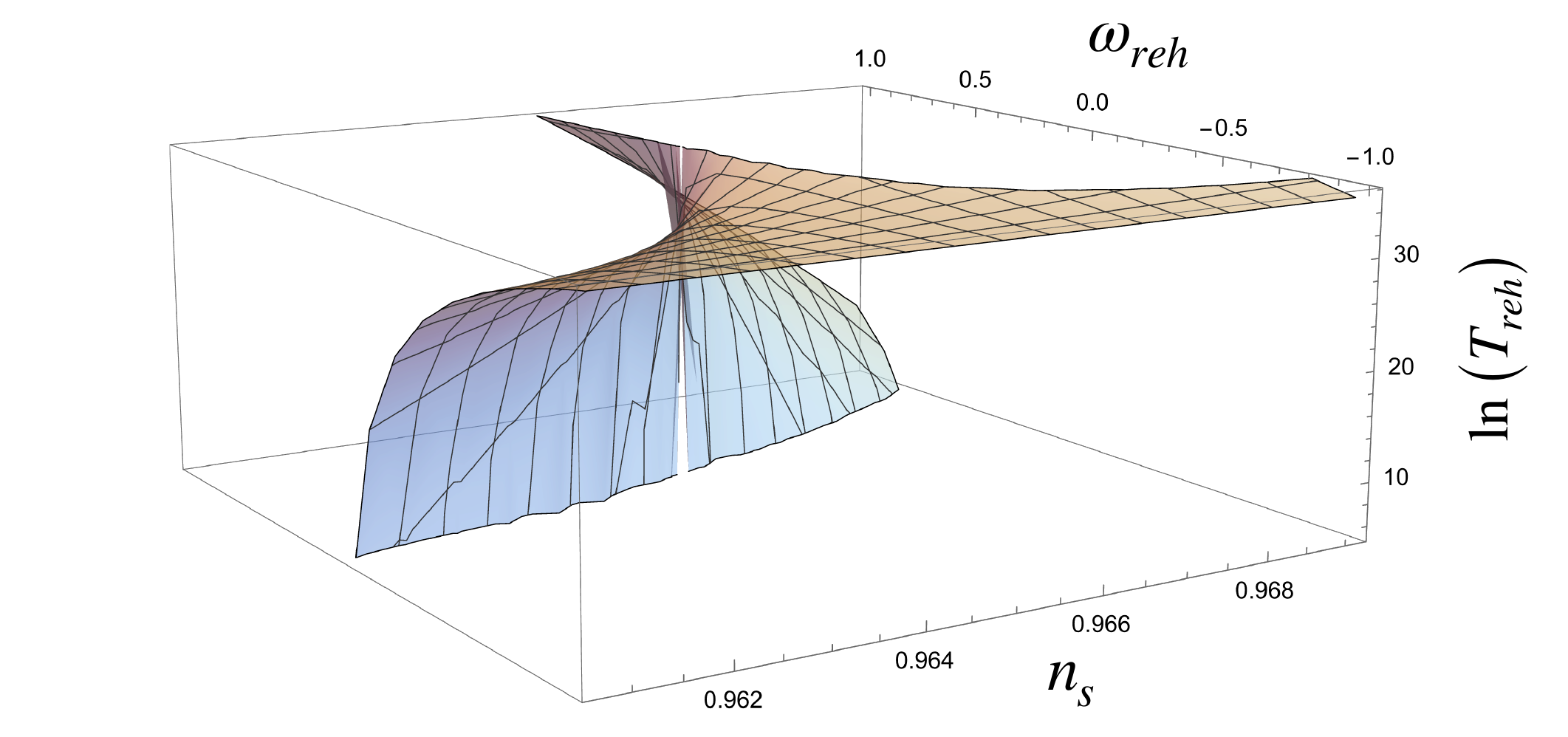}
    \caption{Three-dimensional plot for the reheating temperature $T_\text{reh}$ (logarithmic scale) in terms of  the scalar spectral index $n_s$ and the EoS parameter $\omega_\text{reh}$.
    Here we have used that the tensor to scalar ratio corresponds to the upper bound $r_k=0.039$ and the parameter $\theta=10^{8}M_p^{-2}$ .} 
    \label{Fig_Reh_ns_M3}
\end{figure}

As before, we observe that the instantaneous reheating  
 takes place when $N_\text{reh}\sim 0$ and reheating temperature presents  a maximum valor. Besides, we notice that any temperature between the BBN bound and the  instantaneous temperature is allowed inside the observational bound, 
independently of the value of $\omega_\text{reh}$. In relation to the two plots,
 we note that  from the left panel, the purple and blue curves are more centered around the value within  the Planck's 1$\sigma$, in relation to the right panel, in which we have considered the value of the parameter $\theta=10^{10}M_p^{-2}$. Also, we observe from the right panel that  increases the value of the parameter $\theta$, the model  predicts  a large number of $e-$folds during the reheating epoch ($N_\text{reh}<40$), for the different reheating temperatures. 

In  Fig.\ref{Fig_Reh_ns_M3} we show the three-dimensional plot for the reheating temperature $T_\text{reh}$ (logarithmic scale) in terms of the EoS parameter and the scalar spectral index. From this figure, we note that the maximum value of the reheating temperature or instantaneous reheating  takes place for all values of the EoS parameter $\omega_\text{reh}$  within the Planck's 1$\sigma$ limit. Also, we note that for positive values  of the EoS parameter the reheating temperature can achieve lower values of the reheating temperature within Planck data 1$\sigma$ limit, in comparison when we consider  negative values of the EoS parameter in which the reheating temperatures are higher than $T_\text{reh}>10^{7}$ GeV,  (see also Fig.(\ref{Fig_Reh_M3})).

\section{Concluding remarks}\label{con}

In this article we have studied  the reconstruction of the background variables during the early universe, in the framework  of a generalized Friedmann equation given by the function $\mathcal{F}(H)$ associated  to the Hubble parameter $H$. In this reconstruction we have utilized 
 as cosmological parameter, the scalar spectral index $n_s=n_s(N)$, where $N$ denotes  the number of $e-$folds during the inflationary scenario.

Under the slow roll approximation, we have developed a new general formalism for reconstructing background variables. This involves a new methodology that rewrites the scalar spectral index as a function of the Hubble parameter and its derivatives.
In this new  analysis of this general formalism, we have determined from the attractor given by the scalar spectral index $n_s=n_s(N)$, an integral relation for a first order  differential equation associated to  the Hubble parameter in terms of the number of $e-$folds  $N$, see Eq.(\ref{eqdiff}). Additionally,  we have obtained  a relation between the scalar field and the $e$-folds $N$,  to reconstruct  the  background variables, such as, the Hubble parameter $H=H(\phi)$ and the effective potential $V=V(\phi)$, as  functions of the scalar field $\phi$.

To rebuild the background variables, we have considered the simplest parametrization (attractor) for the scalar spectral index $n_s=n_s(N)$ as a function of the number of $e-$folds $N$ given by $n_s=1-2/N$ for large-$N$. By assuming this parametrization for the scalar spectral index, we have applied  our methodology for three different modified Friedmann equations associated with the various  functions $\mathcal{F}(H)$. Thus, in the reconstruction of the background variables, we have considered the functions;  $\mathcal{F}(H)\propto H^{\beta}$ as a first example, the function $\mathcal{F}(H)=H^2-\gamma H^{-2}$, as a second application and finally the function 
 $\mathcal{F}(H)=H^2+\theta H^4$, as example III. In this sense, using the new methodology for reconstructing the background variables with different functions $\mathcal{F}(H)$, we have found that these reconstructed background variables depend on the sign of the integration constants that arise  from the solutions of the differential equations, which are solved in terms of the number of $e$-folds $N$, for each of the chosen functions $\mathcal{F}(H)$. 
In addition, for the different $\mathcal{F}(H)-$models analyzed, we have constrained these integration constants from the observational parameters, such as the power scalar spectrum and the tensor to scalar ratio, in the particular case in which the number of $e-$folds corresponds to $N=60$. In Figs. \ref{Fig_Potential_M1}, \ref{Fig_Potential_M2} and \ref{Fig_Potential_M3}, we show the evolution of the reconstructed effective potential in terms of the scalar field for the three different $\mathcal{F}(H)-$models studied. In all cases analyzed, we have found that the scalar field begins to roll from the maximum value of the potential (approximately) towards values of the scalar field close to zero, where the number of $e-$folds at the end of inflationary epoch is $N_\text{end}\sim 0$, see e.g., the left panel of Fig.\ref{Fig_Potential_M1}.

In relation to analyze of the reheating era for our different $\mathcal{F}(H)-$models, we have obtained that it is possible to quantify  this epoch in terms of the reheating parameters such as; the temperature, the number of $e-$folds and the EoS parameter during the reheating of the universe. From the reconstruction of the  background variables found during the inflationary epoch from the attractor $n_s=n_s(N)$,  we have been able to obtain these reheating parameters. In particular for the first function studied, we have found that the reheating temperature associated to the power $\beta=1$ and $\beta=2$ present a good compatibility with Planck's 1$\sigma$ bound on the scalar spectral index, for the different values of the EoS parameter $\omega_\text{reh}$, excluding the negative value  $\omega_\text{reh}=-1/3$. For the second function $\mathcal{F}(H)$, we have found that the reheating temperature for the different values of the EoS parameter does not depend of the parameter $\gamma\ll 1$, since the reconstruction of the Hubble parameter and effective potential are not affected strongly for this parameter. In relation to the third function  $\mathcal{F}(H)$, we have obtained that reheating temperature increases when we increase the parameter $\theta$. In addition, we have found that the reheating temperature for positive values of the EoS parameter can achieve lower values in comparison with values of $\omega_\text{reh}<0$.

Regarding  the number of $e-$folds during the reheating for these $\mathcal{F}(H)-$models, we have observed that this number presents a high value when the EoS parameter $\omega_\text{reh}$ is positive (the highest value corresponds to $\omega_\text{reh}=2/3$), see the respective figures associated to the reheating. In the different $\mathcal{F}(H)$-models, we have  found that
 the stage of instantaneous reheating is given when the number of $e-$folds at the end of inflation $N_\text{reh}$ approaches  zero. In the various panels associated to the reheating epoch,  we have determined that this scenario corresponds to the point  in which all lines converge to the value
$N_\text{reh}\simeq 0$.

Finally in this article, we have not addressed the reconstruction of the background variables  for another functions $\mathcal{F}(H)$ from the parametrization on $n_s=n_s(N)$, as well as the reheating study assuming an EoS parameter as a function of the time $\omega_\text{reh}=\omega_\text{reh}(t)$ in  numerical form to obtain the reheating parameters; $T_\text{reh}$ and $N_\text{reh}$, respectively. In this sense, we hope to return to these points in the near future.

\end{document}